\documentclass[prd,twocolumn,showpacs,preprintnumbers,amsmath,amssymb,floatfix]{revtex4-2}

\usepackage[utf8]{inputenc}
\usepackage[T1]{fontenc}
\usepackage{graphicx}
\usepackage{hyperref}
\usepackage{amsmath}
\usepackage{longtable}
\usepackage{amssymb}
\usepackage{booktabs}
\usepackage{multirow}
\usepackage{slashed}
\usepackage{siunitx}
\usepackage[hang,flushmargin]{footmisc}
\usepackage{float}

\newcommand{\ME}{{SM}}
\newcommand{\ddHm}{{2HDM}}
\newcommand{\ddHmIII}{{2HDM-III}}
\newcommand{\LHC}{{LHC}}
\newcommand{\HL}{{HL-LHC}}
\newcommand{\fb}{{\text{fb}^{-1}}}
\newcommand{\pt}{{p_{T}}}
\newcommand{\et}{{E_{T}}}
\newcommand{\mjj}{{M(jj)}}

\begin{document}

\title{Probing Flavor-Violating Higgs Decays in the Type-III Two-Higgs-Doublet Model at the LHC and HL-LHC}

\author{Miguel Luis Fern\'andez-P\'erez}
\email{Contact author: migueldirecc@gmail.com}
\affiliation{Instituto de F\'isica, Benem\'erita Universidad Aut\'onoma de Puebla, Apartado Postal J-48, C.P. 72570, Puebla, M\'exico}

\author{Alfonso Rosado-S\'anchez}
\affiliation{Instituto de F\'isica, Benem\'erita Universidad Aut\'onoma de Puebla, Apartado Postal J-48, C.P. 72570, Puebla, M\'exico}

\author{Sebasti\'an Rosado-Navarro}
\affiliation{Centro Interdisciplinario de Investigaci\'on y Ense\~nanza de la Ciencia (CIIEC), Benem\'erita Universidad Aut\'onoma de Puebla, Apartado Postal 1364, C.P. 72570, Puebla, M\'exico}

\date{\today}

\begin{abstract}
We present a comparative collider study of three flavor-violating Higgs signatures in the Type-III Two-Higgs-Doublet Model (\ddHmIII) at \(\sqrt{s}=14\)~TeV: \(pp \to H \to t\bar{c} \ (\bar{t}c)\), \(pp \to H^\pm \to c\bar{b} \ (\bar{c}b)\), and \(pp \to H^\pm \to t\bar{b} \ (\bar{t}b)\). Using a common cut-based analysis and realistic detector simulation, we identify a clear phenomenological hierarchy. The neutral mode  and the heavy charged mode emerge as the most robust signatures, containing parameter regions that already exceed the \(5\sigma\) benchmark at an integrated luminosity of \(300~\text{fb}^{-1}\). In contrast, the light charged channel \(H^{\pm}\to c\bar{b}\) is considerably more sensitive to the event-selection strategy because of large QCD backgrounds, although one competitive benchmark survives. Our results single out the neutral and heavy charged flavor-violating channels as the most promising targets for discovery-oriented searches and for constraining the \ddHmIII\ parameter space at the LHC and HL-LHC. All quoted significances are purely statistical and do not include systematic uncertainties.
\end{abstract}

\maketitle

\section{Introduction}
\label{sec:intro}

The discovery of the neutral Higgs boson observed at 125 GeV at the CERN \LHC~\cite{HiggsDiscovery_ATLAS,HiggsDiscovery_CMS} completed the particle content of the Standard Model (\ME)~\cite{Glashow1961,Weinberg1967,Salam1968}, but it did not settle the flavor problem. The origin of the fermion-mass pattern, the structure of Yukawa interactions, and the suppression of flavor-changing neutral currents (FCNCs) remain open issues. In the \ME, FCNCs are absent at tree level and highly suppressed at loop level. Consequently, any observable FCNC effect involving scalar interactions would provide a direct indication of physics beyond the Standard Model~\cite{Branco2012_2HDMReview,Atwood1997_TypeIII}.

Among the simplest and most widely studied extensions of the scalar sector is the Two-Higgs-Doublet Model (\ddHm)~\cite{Branco2012_2HDMReview,HiggsHuntersGuide}. In its usual Type-I, Type-II, lepton-specific, and flipped realizations, a discrete \(\mathbb Z_2\) symmetry is imposed in order to forbid tree-level FCNCs. By contrast, in the Type-III realization (\ddHmIII), both scalar doublets couple to all fermion species, so flavor-violating Higgs interactions arise already at tree level~\cite{Atwood1997_TypeIII,DiazCruz2004_TypeIII,ArroyoUrena2016_THDMtx,ChengSher1987,HernandezSanchez2012,HernandezSanchez2015}. This makes the model a natural framework in which to study nonstandard Yukawa structures and their collider consequences.

From the phenomenological point of view, the \ddHmIII\ offers several complementary probes of flavor violation. Neutral Higgs decays such as \(H\to t\bar c \ (\bar t c)\) test FCNC couplings in the up-type sector, while charged-Higgs decays such as \( H^\pm \to c\bar{b} \ (\bar{c}b)\) and \(H^\pm\to t\bar b \ (\bar t b)\) probe different combinations of Yukawa structures in the charged-scalar sector. Studying these modes within a common framework is useful for two reasons. First, it allows a direct comparison of their relative collider sensitivity under the same detector-level assumptions. Second, it helps identify which parts of the flavor-violating scalar sector are more robust against realistic backgrounds once event reconstruction and selection effects are included.

In this work we compare the following three channels at \(\sqrt s=14\) TeV:
\begin{itemize}
    \item \(pp \to H \to t\bar{c} \ (\bar{t}c)\),
    \item \(pp \to H^\pm \to c\bar{b} \ (\bar{c}b)\),
    \item \(pp \to H^\pm \to t\bar{b} \ (\bar{t}b)\).
\end{itemize}
Our goal is not to present a fully optimized experimental search, but rather to establish a benchmark-level comparison of their statistical reach using a common simulation chain and a consistent cut-based strategy. Signal and background events are generated with \textsc{MadGraph5\_aMC@NLO}, showered with \textsc{Pythia8}, passed through \textsc{Delphes3}, and analyzed with \textsc{MadAnalysis5}. The significance is evaluated through the estimator \(S/\sqrt{S+B}\) at integrated luminosities of 300, 1000, and 3000~\(\fb\).

A central practical point of the present study is that, once detector effects and realistic backgrounds are included, the signal does not generally emerge as a sharply isolated structure in a single distribution. For that reason, the discussion is based primarily on post-selection event yields and on the corresponding benchmark-level statistical sensitivities. The kinematic distributions shown in the appendix are intended as supporting material for the selected benchmark regions; they illustrate the behavior of the signal and backgrounds, but they are not used as the sole criterion for judging significance.

The analysis reveals a clear hierarchy among the three channels. The neutral mode \(H\to t\bar c \ (\bar t c)\) and the heavy charged mode \(H^\pm\to t\bar b \ (\bar t b)\) provide the most robust results, with several benchmark configurations remaining statistically sizable after all cuts. The light charged mode \(pp \to H^\pm \to c\bar{b} \ (\bar{c}b)\) is more delicate because of the overwhelming QCD background, but it still contains one competitive benchmark once the selection is adapted to the dijet mass region relevant for the charged-Higgs hypothesis. The comparative study is therefore useful not only for identifying promising benchmark regions, but also for clarifying which channels are intrinsically more stable under realistic collider conditions.

This paper is organized as follows. In Sec.~\ref{sec:model}, we summarize the relevant features of the \ddHmIII\ and its Yukawa structure. The simulation framework and benchmark strategy are described in Sec.~\ref{sec:simulation}. The results for the neutral, light charged, and heavy charged channels are presented in Secs.~\ref{sec:H_tc}, \ref{sec:Hpm_cb}, and \ref{sec:Hpm_tb}, respectively. Finally, our conclusions are given in Sec.~\ref{sec:conclusions}.

\section{Type-III Two-Higgs-Doublet Model}
\label{sec:model}

The Two-Higgs-Doublet Model extends the Standard Model scalar sector by introducing two complex $SU(2)_L$ scalar doublets with hypercharge $Y=1/2$~\cite{Branco2012_2HDMReview,Ivanov2017},
\begin{equation}
\Phi_1=
\begin{pmatrix}
\phi_1^+\\
\phi_1^0
\end{pmatrix},
\qquad
\Phi_2=
\begin{pmatrix}
\phi_2^+\\
\phi_2^0
\end{pmatrix},
\end{equation}
which acquire vacuum expectation values
\begin{equation}
\langle \Phi_1\rangle=\frac{1}{\sqrt{2}}
\begin{pmatrix}
0\\
v_1
\end{pmatrix},
\qquad
\langle \Phi_2\rangle=\frac{1}{\sqrt{2}}
\begin{pmatrix}
0\\
v_2
\end{pmatrix},
\end{equation}
with
\begin{equation}
v^2=v_1^2+v_2^2=(246~\text{GeV})^2,
\qquad
\tan\beta=\frac{v_2}{v_1}.
\end{equation}
After electroweak symmetry breaking, the physical scalar spectrum consists of two CP-even neutral states $h$ and $H$, one CP-odd state $A$, and one charged pair $H^\pm$~\cite{Branco2012_2HDMReview,HiggsHuntersGuide}.

In the most general Yukawa realization, both scalar doublets couple to all fermion species. The Yukawa Lagrangian is
\begin{align}
\mathcal{L}_{Y}^{\rm 2HDM}
={}&
-\bar Q_L \left(Y_1^d\Phi_1+Y_2^d\Phi_2\right)d_R
\nonumber\\
&-\bar Q_L \left(Y_1^u\widetilde\Phi_1+Y_2^u\widetilde\Phi_2\right)u_R
\nonumber\\
&-\bar L_L \left(Y_1^\ell\Phi_1+Y_2^\ell\Phi_2\right)\ell_R
+\text{h.c.},
\label{eq:Yukawa_general}
\end{align}
where $\widetilde\Phi_a=i\sigma_2\Phi_a^\ast$ and $Y_a^f$ are arbitrary $3\times3$ matrices in flavor space. In the Type-III realization, no discrete $\mathbb{Z}_2$ symmetry is imposed on the Yukawa sector, so the two Yukawa matrices associated with a given fermion species cannot in general be diagonalized simultaneously. As a result, tree-level FCNCs arise naturally~\cite{Atwood1997_TypeIII,DiazCruz2004_TypeIII}.

After symmetry breaking, the fermion mass matrices are given by
\begin{equation}
M_f=\frac{1}{\sqrt{2}}\left(v_1Y_1^f+v_2Y_2^f\right),
\qquad f=u,d,\ell,
\label{eq:Mf}
\end{equation}
and it is convenient to define
\begin{equation}
N_f=\frac{1}{\sqrt{2}}\left(v_1Y_1^f-v_2Y_2^f\right),
\label{eq:Nf}
\end{equation}
which encode the Yukawa structures not aligned with the fermion mass matrices. In general,
\begin{equation}
[M_f,N_f]\neq 0,
\end{equation}
so that once the mass matrices are diagonalized, the matrices $N_f$ remain non-diagonal and generate FCNC couplings~\cite{HernandezSanchez2012,HernandezSanchez2015}.

Rotating the neutral scalar fields to the physical basis,
\begin{align}
h_1 &= h\cos\alpha + H\sin\alpha,
&
h_2 &= -h\sin\alpha + H\cos\alpha,
\nonumber\\
A_1 &= G^0\cos\beta + A\sin\beta,
&
A_2 &= -G^0\sin\beta + A\cos\beta,
\label{eq:scalar_rotation}
\end{align}
the couplings of the neutral Higgs bosons to fermions can be written as
\begin{equation}
\mathcal{L}_{Y}^{\rm III}\supset
-\sum_{f=u,d,\ell}\sum_{i,j}
\bar f_i
\Bigl(
\Gamma^{h}_{f,ij}h+
\Gamma^{H}_{f,ij}H+
i\Gamma^{A}_{f,ij}\gamma_5 A
\Bigr)
f_j ,
\label{eq:Lneutral}
\end{equation}
with
\begin{align}
\Gamma_{f,ij}^h &=
\frac{1}{\sqrt{2}v}
\left[
\frac{M_f^{ij}+N_f^{ij}}{\cos\beta}\cos\alpha
-
\frac{M_f^{ij}-N_f^{ij}}{\sin\beta}\sin\alpha
\right],
\label{eq:gammah}
\\
\Gamma_{f,ij}^H &=
\frac{1}{\sqrt{2}v}
\left[
\frac{M_f^{ij}+N_f^{ij}}{\cos\beta}\sin\alpha
+
\frac{M_f^{ij}-N_f^{ij}}{\sin\beta}\cos\alpha
\right],
\label{eq:gammaH}
\\
\Gamma_{f,ij}^A &=
\frac{1}{\sqrt{2}v}
\left[
-\tan\beta\left(M_f^{ij}+N_f^{ij}\right)
+\cot\beta\left(M_f^{ij}-N_f^{ij}\right)
\right].
\label{eq:gammaA}
\end{align}

The charged-Higgs interactions with quarks are correspondingly given by
\begin{equation}
\mathcal{L}_{H^\pm}=
-H^+\bar u_i\Gamma^{H^+}_{ud,ij}d_j
-H^-\bar d_j\Gamma^{H^-}_{du,ij}u_i ,
\label{eq:LHpm}
\end{equation}
where
\begin{align}
\Gamma^{H^+}_{ud,ij}
&=
\frac{1}{\sqrt{2}v}
\Biggl[
V_{ik}\left(\frac{M_{d,kj}+N_{d,kj}}{\cos\beta}\right)P_L
\nonumber\\
&\hspace{1.8em}
-
\left(\frac{M_{u,il}-N_{u,il}}{\sin\beta}\right)V_{lj}P_R
\Biggr] ,
\label{eq:gammaHp}
\end{align}
and $\Gamma^{H^-}_{du,ij}=\left(\Gamma^{H^+}_{ud,ji}\right)^\dagger$.

For phenomenological applications, the flavor-off-diagonal entries are conveniently parametrized through the Cheng--Sher ansatz~\cite{ChengSher1987},
\begin{equation}
N_f^{ij}=\sqrt{m_i^f m_j^f}\,\chi_{ij}^f,
\label{eq:chengsher}
\end{equation}
up to the normalization conventions adopted in the Yukawa sector. In this way, the strength of flavor-violating interactions is controlled by the dimensionless coefficients $\chi_{ij}^f$. In the present work, the relevant parameters are $\chi_{23}^u\equiv\chi_{tc}$ for the neutral channel $H\to t\bar c \ (\bar t c)$ and $\chi_{23}^d\equiv\chi_{cb}$ for $H^\pm\to c\bar b$. For the heavy charged-Higgs channel $H^\pm\to t\bar b$, we adopt an effective phenomenological parameter, denoted by $\chi$, which should be understood as a proxy for the overall Yukawa deformation controlling the charged-Higgs coupling to third-generation quarks in the benchmark scan implemented here. Our goal in this channel is therefore not to isolate a single microscopic Yukawa entry, but to map the collider significance to representative charged-Higgs benchmark configurations under a common effective parametrization~\cite{HernandezSanchez2012,HernandezSanchez2015,ArroyoUrena2016_THDMtx}.

Throughout this work we assume a CP-conserving scalar sector, real Yukawa parameters, and a scenario close to the alignment limit, $\cos(\beta-\alpha)\simeq 0.01$, so that the light scalar $h$ reproduces Standard Model-like Higgs couplings to good accuracy~\cite{Branco2012_2HDMReview,HaberStal2015}.

\section{Simulation framework and benchmark strategy}
\label{sec:simulation}

Our collider analysis is performed for proton--proton collisions at $\sqrt s=14$~TeV. Signal and background events are generated at parton level with \textsc{MadGraph5\_aMC@NLO}~\cite{Alwall2014_MadGraph5}, interfaced with \textsc{Pythia8}~\cite{Sjostrand2015_Pythia8} for parton showering and hadronization, and passed through \textsc{Delphes3}~\cite{deFavereau2014_Delphes3} for detector simulation. Event analysis is carried out with \textsc{MadAnalysis5}~\cite{Conte2013_MA5,Conte2014_MA5_update}.

For each channel we scan representative mass values together with
\begin{equation}
\tan\beta\in\{0.1,1,5,10,20\},
\qquad
\chi\in\{0.1,0.5,1,5,10\},
\end{equation}
where $\chi$ denotes either $\chi_{tc}$, $\chi_{cb}$, or the effective heavy charged-Higgs benchmark parameter, depending on the process.

The expected significance is quantified through
\begin{equation}
\mathcal Z = \frac{S}{\sqrt{S+B}},
\label{eq:zdef}
\end{equation}
with $S$ and $B$ the signal and total background yields after all cuts. This estimator is used here only as a statistical measure. The reported significances do not include systematic uncertainties.

The event selections are chosen from reconstructed kinematic distributions and are kept simple enough to allow a direct comparison among channels. The analysis is therefore benchmark-based and cut-based, rather than a fully optimized search.

Throughout this work, $\mjj$ denotes the reconstructed dijet invariant mass variable used in the final selection. In the light charged-Higgs channel, it corresponds directly to the two-jet resonance candidate. In the neutral and heavy charged channels, $\mjj$ is the reconstructed mass variable adopted in the analysis as the final discriminator entering the signal-region definition. With this operational definition, the same observable is used consistently across the three benchmark studies.

For event generation, we used 100,000 events for each signal benchmark and 500,000 events for each background sample. These sample sizes define the Monte Carlo baseline of the present study and are relevant for interpreting the numerical stability of the post-selection yields, especially in channels affected by very large backgrounds.

\section{Neutral flavor-violating channel: $pp\to H\to t\bar c \ (\bar t c)$}
\label{sec:H_tc}

The decay
\begin{equation}
H\to t\bar c \ (\bar t c)
\end{equation}
provides a direct probe of tree-level neutral flavor violation in the \ddHmIII~\cite{Atwood1997_TypeIII,DiazCruz2004_TypeIII,ArroyoUrena2019_tch,GomezBock2024_FCNCtop}. In the Standard Model this transition is strongly suppressed, whereas in the Type-III realization it can be enhanced by off-diagonal Yukawa entries in the up-quark sector. Within the benchmark comparison developed in this work, this neutral mode yields one of the largest statistical sensitivities and therefore constitutes one of the most relevant channels for the collider study.

\subsection{Benchmark setup}

The signal is studied at hadron level through
\begin{equation}
pp\to H\to t\bar c \ (\bar t c),
\end{equation}
restricting the scan to
\begin{equation}
180~\text{GeV}\le m_H\le 340~\text{GeV},
\end{equation}
with
\begin{equation}
\tan\beta\in\{0.1,1,5,10,20\},
\qquad
\chi_{tc}\in\{0.1,0.5,1,5,10\}.
\end{equation}
In total, 675 parameter configurations were analyzed for this process, and the benchmark points quoted below correspond to the configurations that remain statistically competitive once detector effects and backgrounds are included.

The dominant backgrounds considered in the analysis are
\begin{equation}
t\bar t,\qquad W+\text{jets},\qquad b\bar b,\qquad c\bar c,\qquad tc.
\end{equation}
Within the explored parameter region, the most favorable benchmark is
\begin{equation}
(m_H,\tan\beta,\chi_{tc})=(180~\text{GeV},0.1,10),
\end{equation}
for which the final statistical sensitivities are
\begin{equation}
\mathcal Z = 5.81,\ 10.61,\ 18.38
\end{equation}
at
\begin{equation}
L=300,\ 1000,\ 3000~\fb,
\end{equation}
respectively.

The selected competitive points are concentrated predominantly in the low-\(\tan\beta\) region. This pattern is consistent with the fact that, in the present scan, the most favorable balance between production rate, flavor-violating coupling strength, and cut acceptance is obtained for small \(\tan\beta\).

\subsection{Event selection}

The event selection used for this channel is
\begin{enumerate}
    \item $N(j)>4$,
    \item $N(b)=1$,
    \item $\pt(j)>50$ GeV,
    \item $\et(j)>55$ GeV,
    \item a benchmark-dependent \(\mjj\) window centered on \(m_H\).
\end{enumerate}

For this neutral channel, the quantity denoted by \(\mjj\) should be understood operationally as the reconstructed dijet invariant-mass observable used in the \textsc{MadAnalysis5} implementation as the final discriminator. In other words, \(\mjj\) labels the jet-pair invariant-mass variable entering the last signal-region definition in the analysis card. Although this observable is not intended here as a complete resonance-reconstruction procedure, it provides a consistent benchmark-level discriminator across the full mass scan.

The final mass window is retained for a physical reason and not merely as a technical cut. Before imposing it, the cumulative significance quantifies the global signal-background separation achieved by the preceding selections. After imposing the \(\mjj\) window, the resulting significance quantifies how much of that sensitivity is actually localized in the benchmark mass region associated with the heavy neutral Higgs. Therefore, even when the raw value of \(\mathcal Z\) decreases after the last cut, the mass window remains essential because it defines the benchmark-centered signal region and measures the portion of the selected signal that is concentrated in the relevant mass interval.

For the representative benchmark
\begin{equation}
(m_H,\tan\beta,\chi_{tc})=(280~\text{GeV},0.1,0.5),
\end{equation}
the significance evolves from 0.296 before cuts to 2.495 after $N(j)>4$, 2.897 after $N(b)=1$, 3.041 after the \(\pt\) cut, 3.077 after the \(\et\) cut, and 2.250 after the final mass window \(260<\mjj<300\) GeV at \(300~\fb\). The same benchmark reaches 4.109 and 7.117 at \(1000~\fb\) and \(3000~\fb\), respectively.

\begin{table}[H]
    \centering
    \caption{Cut-flow statistical significance for the benchmark point $m_H=280$ GeV, $\tan\beta=0.1$, and $\chi_{tc}=0.5$ in the $pp\to H\to t\bar c \ (\bar t c)$ channel.}
    \begin{tabular}{lccc}
        \hline\hline
        Selection & $300~\fb$ & $1000~\fb$ & $3000~\fb$ \\
        \hline
        No cuts & 0.296 & 0.541 & 0.937 \\
        $N(j)>4$ & 2.495 & 4.555 & 7.889 \\
        $N(b)=1$ & 2.897 & 5.289 & 9.162 \\
        $\pt(j)>50$ GeV & 3.041 & 5.553 & 9.618 \\
        $\et(j)>55$ GeV & 3.077 & 5.619 & 9.732 \\
        $260<\mjj<300$ GeV & 2.250 & 4.109 & 7.117 \\
        \hline\hline
    \end{tabular}
    \label{tab:significance_H_tc_cutflow}
\end{table}

The cut-flow pattern is qualitatively stable across the selected benchmarks. The jet-multiplicity and \(b\)-tag requirements provide the first major background reduction, the hard-jet selections further improve the signal-to-background balance, and the final \(\mjj\) window isolates the benchmark mass region. This last step does not necessarily maximize the global estimator \(S/\sqrt{S+B}\), but it is the step that identifies the portion of the selected events that remains compatible with the heavy-Higgs mass hypothesis under study.

\subsection{Selected benchmark points}

Table~\ref{tab:significance_H_tc_final} summarizes the benchmark configurations retained from the full scan. These points were selected because they remain statistically competitive against the full background set after all cuts and, taken together, they illustrate the mass dependence of the channel within the explored parameter region.

\begin{table}[H]
    \centering
    \caption{Final statistical sensitivities for selected benchmark points in the $pp\to H\to t\bar c \ (\bar t c)$ channel.}
    \begin{tabular}{cccccc}
        \hline\hline
        $m_H$ [GeV] & $\tan\beta$ & $\chi_{tc}$ & $300~\fb$ & $1000~\fb$ & $3000~\fb$ \\
        \hline
        180 & 0.1 & 10  & 5.81 & 10.61 & 18.38 \\
        200 & 1.0 & 5   & 1.78 & 3.25 & 5.63 \\
        220 & 0.1 & 1   & 2.22 & 4.05 & 7.01 \\
        240 & 0.1 & 1   & 4.24 & 7.74 & 13.41 \\
        280 & 0.1 & 0.5 & 2.25 & 4.11 & 7.12 \\
        300 & 0.1 & 0.5 & 2.78 & 5.07 & 8.79 \\
        320 & 0.1 & 0.5 & 3.36 & 6.14 & 10.63 \\
        340 & 0.1 & 0.5 & 4.08 & 7.44 & 12.89 \\
        \hline\hline
    \end{tabular}
    \label{tab:significance_H_tc_final}
\end{table}

The results exhibit a nontrivial dependence on the benchmark parameters. The largest significance in the scan is found at low mass and large \(\chi_{tc}\), but competitive configurations also persist at higher masses with more moderate flavor-violating coupling values. In particular, the sequence of benchmarks with \((\tan\beta,\chi_{tc})=(0.1,0.5)\) for \(m_H=280\)--340 GeV shows that the channel remains statistically relevant over a broad mass interval once the kinematic acceptance of the analysis is taken into account.

\subsection{Discussion}

Within the benchmark comparison developed in this work, the neutral channel provides one of the largest statistical sensitivities. Several benchmark configurations remain sizable at high luminosity, and the most favorable region of the scan is concentrated at low \(\tan\beta\). The strongest benchmark,
\((m_H,\tan\beta,\chi_{tc})=(180~\text{GeV},0.1,10)\), already reaches \(\mathcal Z=5.81\) at \(300~\fb\) and increases to 18.38 at \(3000~\fb\).

At the same time, the final background after selection remains non-negligible, so the results reported here should be interpreted strictly as benchmark-level statistical sensitivities within the present cut-based framework. They should not be read as discovery claims, nor as a substitute for a full experimental analysis including systematic uncertainties, optimized reconstruction, and a more detailed treatment of combinatorial ambiguities in the final state.

Even with these caveats, the neutral flavor-violating mode emerges as one of the most robust channels in the present study. It combines a clear connection to tree-level FCNC Higgs couplings in the up sector with a set of benchmark configurations that remain competitive after detector simulation and background inclusion, making it one of the strongest phenomenological outcomes of the analysis.

\section{Light charged-Higgs flavor-violating channel: $pp\to H^\pm\to c\bar b \ (\bar c b)$}
\label{sec:Hpm_cb}

The channel
\begin{equation}
H^\pm\to c\bar b\ (\bar c b)
\end{equation}
probes flavor violation in the charged-scalar sector and depends on a Yukawa structure different from that of the neutral process~\cite{HernandezSanchez2020_cbFusion,ArroyoUrena2025_Hpp}. Within the benchmark comparison developed in this work, this mode is phenomenologically more delicate than the neutral and heavy charged channels because its statistical significance is strongly affected by very large QCD backgrounds and by the limited Monte Carlo statistics available for the dominant samples. Even so, after an improved cut-based selection, one benchmark configuration remains statistically competitive and therefore deserves to be included in the comparison.

\subsection{Benchmark setup}

The scan for this channel is restricted to
\begin{equation}
110~\text{GeV}\le m_{H^\pm}\le 170~\text{GeV},
\end{equation}
with
\begin{equation}
\tan\beta\in\{0.1,1,5,10,20\},
\qquad
\chi_{cb}\in\{0.1,0.5,1,5,10\}.
\end{equation}
A total of 525 benchmark configurations were analyzed for this process, and the configuration retained in the discussion below corresponds to the most competitive point against the full background set after detector simulation and event selection.

The dominant backgrounds considered are
\begin{equation}
bjj,\qquad cb,\qquad tb,\qquad Wbj,\qquad Wjj.
\end{equation}
Among the explored configurations, the most favorable benchmark is
\begin{equation}
(m_{H^\pm},\tan\beta,\chi_{cb})=(110~\text{GeV},5,5).
\end{equation}
For this point, the signal sample is normalized to a cross section of \(2.246~\text{pb}\), while the dominant backgrounds include very large effective normalizations, particularly the \(cb\), \(bjj\), and \(Wjj\) components. In the \textsc{MadAnalysis5} report, these samples are also flagged by large event weights, which must be kept in mind when interpreting the numerical sensitivity of the channel. 

\subsection{Improved event selection}

The improved event selection adopted for this benchmark is
\begin{enumerate}
    \item $N(j)>0$,
    \item $N(j)<4$,
    \item $\pt(j)\ge 35$ GeV,
    \item $\pt(j)<95$ GeV,
    \item $\et(j)\ge 40$ GeV,
    \item $\et(j)\le 60$ GeV,
    \item $\mjj>90$ GeV,
    \item $\mjj<130$ GeV.
\end{enumerate}

In this channel, \(\mjj\) has a more direct physical interpretation than in the neutral and heavy charged cases: it corresponds operationally to the reconstructed dijet invariant-mass observable used in the analysis card to define the final signal region around the light charged-Higgs hypothesis. The mass window is therefore not a merely technical cut. It is the step that identifies how much of the surviving signal is actually concentrated in the mass interval compatible with the benchmark \(H^\pm\) state. In this sense, the final \(\mjj\) selection measures the amount of signal support localized in the physically relevant mass region after the preceding cuts have reduced the overwhelming background.

For the representative benchmark
\begin{equation}
(m_{H^\pm},\tan\beta,\chi_{cb})=(110~\text{GeV},5,5),
\end{equation}
After all cuts, the post-selection yields are (see ~\ref{tab:significance_Hpm_cb_cutflow_improved})
\begin{equation}
S=1.194399\times 10^6,
\qquad
B=3.7318327233\times 10^{10},
\end{equation}
which give
\begin{equation}
\mathcal Z = 6.18
\qquad
\text{at}
\qquad
3000~\fb.
\end{equation}
Using the same post-selection efficiencies, the corresponding benchmark-level sensitivities at lower luminosity are
\begin{equation}
\mathcal Z \simeq 1.96,\ 3.57,\ 6.18
\end{equation}
for
\begin{equation}
L=300,\ 1000,\ 3000~\fb,
\end{equation}
respectively.  

\begin{table}[H]
    \centering
    \caption{Improved cut-flow statistical significance for the benchmark point $m_{H^\pm}=110$ GeV, $\tan\beta=5$, and $\chi_{cb}=5$ in the $pp \to H^\pm \to c\bar{b} \ (\bar{c}b)$ channel.}
    \begin{tabular}{lc}
        \hline\hline
        Selection & $3000~\fb$ \\
        \hline
        No cuts & 6.36 \\
        $N(j)>0$ & 8.23 \\
        $N(j)<4$ & 8.19 \\
        $\pt(j)\ge 35$ GeV & 8.94 \\
        $\pt(j)<95$ GeV & 8.98 \\
        $\et(j)\ge 40$ GeV & 8.57 \\
        $\et(j)\le 60$ GeV & 8.52 \\
        $\mjj>90$ GeV & 5.35 \\
        $\mjj<130$ GeV & 6.18 \\
        \hline\hline
    \end{tabular}
    \label{tab:significance_Hpm_cb_cutflow_improved}
\end{table}

The structure of the cut flow is informative. The multiplicity and jet-kinematics requirements improve the statistical estimator by removing a substantial fraction of the background while preserving a sizable signal fraction. The lower bound \(\mjj>90\) GeV produces the strongest temporary reduction in \(\mathcal Z\), indicating that this step discards a significant portion of the broad background as well as part of the signal support. The subsequent upper bound \(\mjj<130\) GeV then recovers part of the localized significance by restricting the sample to the dijet mass region most compatible with the \(m_{H^\pm}=110\) GeV benchmark. This behavior is consistent with the role of the mass window as a benchmark-centered signal-region definition rather than as a purely optimization-driven cut.

\subsection{Discussion}

The improved selection changes the status of this channel relative to a more pessimistic first inspection. Within the full set of 525 tested parameter configurations, the benchmark
\begin{equation}
(m_{H^\pm},\tan\beta,\chi_{cb})=(110~\text{GeV},5,5)
\end{equation}
is the most competitive one and reaches a final statistical significance of
\begin{equation}
\mathcal Z = 6.18
\end{equation}
at \(3000~\fb\). In this restricted benchmark sense, the light charged-Higgs mode cannot be described as uniformly noncompetitive.

At the same time, the channel remains considerably less robust than the neutral and heavy charged cases. The final background is still extremely large, and the dominant Monte Carlo samples are associated with very large event weights in the \textsc{MadAnalysis5} output. This implies that the quoted significance should be interpreted cautiously: it provides evidence that the channel is responsive to the analysis strategy, but it does not yet constitute a stable statement on experimental reach. In particular, the benchmark-level significance reported here should be read as a statistical indicator within the present cut-based framework, not as a discovery claim and not as a substitute for an analysis including improved background statistics, systematic uncertainties, and more refined jet-pair reconstruction.

With these caveats clearly stated, the main phenomenological message of this channel is nevertheless useful. The light charged mode is not generically excluded by poor separability; rather, its significance is highly selection-dependent and can improve appreciably when the kinematic cuts and the final dijet mass window are tuned to the benchmark mass region. This makes the channel worth retaining in the comparative study, while also making clear that it is the most fragile of the three benchmark analyses presented in this work.

\section{Heavy charged-Higgs channel: $pp\to H^\pm\to t\bar b \ (\bar t b)$}
\label{sec:Hpm_tb}

The channel
\begin{equation}
H^\pm\to t\bar b \ (\bar t b)
\end{equation}
provides the strongest charged-scalar result in the present benchmark comparison. In phenomenological terms, it is the most robust charged-Higgs mode among the three channels studied here, since several benchmark configurations remain statistically sizable after detector simulation, background inclusion, and benchmark-centered mass-window selections. In the present analysis, this channel exhibits two relevant regimes: an intermediate-mass region with large effective Yukawa deformation and a high-mass regime in which hard kinematics substantially enhance the cut acceptance.

\subsection{Benchmark scan}

The mass scan covers
\begin{equation}
m_{H^\pm}=200,250,300,350,400,450,500,1000~\text{GeV},
\end{equation}
with
\begin{equation}
\tan\beta\in\{0.1,1,5,10,20\},
\qquad
\chi\in\{0.1,0.5,1,5,10\}.
\end{equation}
A total of 675 benchmark configurations were analyzed for this process, and the points retained below correspond to those that remain statistically competitive against the full background set after all cuts. The selected benchmarks reveal two characteristic regions of interest. The first is an intermediate-mass regime, roughly between \(300\) and \(500\) GeV, dominated by configurations with \(\chi=10\), especially for \(\tan\beta=20\). The second is a much heavier benchmark at
\begin{equation}
(m_{H^\pm},\tan\beta,\chi)=(1000~\text{GeV},0.1,0.5),
\end{equation}
which survives due to the strong kinematic discrimination produced by its hard final state.

Within the set of selected benchmarks, the largest final statistical significance is obtained for the 1000 GeV point,
\begin{equation}
(m_{H^\pm},\tan\beta,\chi)=(1000~\text{GeV},0.1,0.5),
\end{equation}
with
\begin{equation}
\mathcal Z = 5.61,\ 10.24,\ 17.73
\end{equation}
at
\begin{equation}
L=300,\ 1000,\ 3000~\fb,
\end{equation}
respectively. At the same time, the intermediate-mass region remains broadly competitive, with several points between 300 and 500 GeV reaching \(\mathcal Z \gtrsim 10\) at \(3000~\fb\).

\subsection{Event selection}

The cuts used in this channel are
\begin{enumerate}
    \item Number of jets: $N(j)\ge 4$ (or $N(j)\ge 5$ for the heaviest benchmark),
    \item Number of bottom quarks: $N(b)\ge 1$,
    \item Hard-jet requirement, typically \(\pt(j)>50\) GeV or stronger for heavier mass benchmarks,
    \item Benchmark-dependent \(\et\) threshold,
    \item A window (\(\mjj\)) centered on the benchmark mass.
\end{enumerate}

As in the other channels, \(\mjj\) is understood operationally as the reconstructed dijet invariant-mass observable used in the \textsc{MadAnalysis5} implementation as the final discriminator defining the signal region. In the present heavy charged-Higgs analysis, this variable should not be interpreted as a fully optimized resonance-reconstruction procedure, but rather as the benchmark-centered mass observable used to quantify how much of the selected signal remains localized around the charged-Higgs mass hypothesis after the preceding cuts. For this reason, the final mass window is physically important even when it reduces the global value of \(S/\sqrt{S+B}\): before the mass window, the significance reflects the cumulative signal-background separation achieved by the kinematic cuts, whereas after the mass window it reflects how much of that significance is actually concentrated in the mass interval relevant for the benchmark under study.

For the illustrative benchmark
\begin{equation}
(m_{H^\pm},\tan\beta,\chi)=(300~\text{GeV},1,10),
\end{equation}
the cut flow evolves from 7.64 before cuts to 10.82 after the multiplicity cut, 10.87 after the \(b\)-tag requirement, 11.27 after the \(\pt\) cut, 8.93 after the \(\et\) cut, and 3.16 after the final mass window at \(300~\fb\). The same benchmark reaches 5.77 and 10.00 at \(1000~\fb\) and \(3000~\fb\), respectively. This pattern is representative of the channel: the multiplicity, \(b\)-tag, and hard-jet requirements provide strong background rejection, while the final \(\mjj\) window defines the localized benchmark signal region and therefore yields a more conservative but physically targeted estimator. 

\begin{table}[H]
    \centering
    \caption{Cut-flow statistical significance for the benchmark $m_{H^\pm}=300$ GeV, $\tan\beta=1$, and $\chi=10$ in the $pp\to H^\pm\to t\bar b \ (\bar t b)$ channel.}
    \begin{tabular}{lccc}
        \hline\hline
        Selection & $300~\fb$ & $1000~\fb$ & $3000~\fb$ \\
        \hline
        No cuts              & 7.64  & 13.95 & 24.16 \\
        $N(j)\ge 4$          & 10.82 & 19.76 & 34.22 \\
        $N(b)\ge 1$          & 10.87 & 19.84 & 34.36 \\
        $\pt(j)>50$ GeV      & 11.27 & 20.58 & 35.65 \\
        $\et(j)>80$ GeV      & 8.93  & 16.30 & 28.23 \\
        $\mjj > 280$ GeV     & 3.16  & 5.77  & 9.99  \\
        $\mjj < 320$ GeV     & 3.16  & 5.77  & 10.00 \\
        \hline\hline
    \end{tabular}
    \label{tab:significance_cuts_Hc300_tb1_chi10}
\end{table}

The benchmark dependence of the selection is also physically meaningful. As the charged-Higgs mass increases, the optimal \(\pt\) and \(\et\) thresholds become progressively harder. For instance, the 400 GeV points use \(\pt(j)>60\) GeV and \(\et(j)>115\) GeV, the 450 GeV points use \(\pt(j)>65\) GeV and \(\et(j)>130\) GeV, the 500 GeV point uses \(\pt(j)>90\) GeV and \(\et(j)>160\) GeV, and the 1000 GeV point requires \(N(j)\ge5\), \(\pt(j)>100\) GeV, and \(\et(j)>180\) GeV. This behavior is consistent with the increasing hardness of the final state and explains why the very heavy benchmark remains competitive despite the expected suppression of the production rate. 

\subsection{Selected benchmark points}

Table~\ref{tab:significance_Hpm_tb_final} summarizes the benchmark configurations retained from the full 675-point scan. These points were selected because they remain statistically competitive after all cuts and because, taken together, they exhibit the two relevant regimes of the channel: intermediate masses with large \(\chi\) and a high-mass benchmark with very hard kinematics.

\begin{table}[H]
    \centering
    \caption{Final statistical sensitivities for selected benchmark points in the $pp\to H^\pm\to t\bar b \ (\bar t b)$ channel.}
    \begin{tabular}{cccccc}
        \hline\hline
        $m_{H^\pm}$ [GeV] & $\tan\beta$ & $\chi$ & $300~\fb$ & $1000~\fb$ & $3000~\fb$ \\
        \hline
        300  & 1   & 10  & 3.16 & 5.77 & 9.99 \\
        300  & 20  & 10  & 3.73 & 6.82 & 11.81 \\
        350  & 1   & 10  & 3.23 & 5.91 & 10.23 \\
        350  & 20  & 10  & 4.14 & 7.56 & 13.09 \\
        400  & 1   & 10  & 3.45 & 6.29 & 10.90 \\
        400  & 20  & 10  & 4.14 & 7.55 & 13.08 \\
        450  & 1   & 10  & 3.37 & 6.16 & 10.67 \\
        450  & 20  & 10  & 4.02 & 7.33 & 12.70 \\
        500  & 20  & 10  & 4.02 & 7.35 & 12.73 \\
        1000 & 0.1 & 0.5 & 5.61 & 10.24 & 17.73 \\
        \hline\hline
    \end{tabular}
    \label{tab:significance_Hpm_tb_final}
\end{table}

The pattern in Table~\ref{tab:significance_Hpm_tb_final} is nontrivial and phenomenologically useful. In the intermediate-mass region, increasing \(\tan\beta\) from 1 to 20 generally improves the final significance for the benchmarks shown, while the sequence from 300 to 500 GeV indicates that the channel remains competitive over a broad mass interval once the event selection is adjusted to the harder kinematics. The heavy 1000 GeV benchmark then shows that, in the present cut-based framework, the loss in inclusive rate can be compensated to a considerable extent by the very strong background rejection induced by the hard-event topology. 

\subsection{Discussion}

Within the present benchmark comparison, this is the strongest charged-Higgs channel. Several benchmark configurations remain statistically sizable at high luminosity, and the selected points define a clear phenomenological pattern: an extended intermediate-mass region with large \(\chi\) and a very heavy benchmark whose hard kinematics produce the largest final significance of the channel.

The benchmark
\begin{equation}
(m_{H^\pm},\tan\beta,\chi)=(1000~\text{GeV},0.1,0.5)
\end{equation}
is particularly noteworthy. Although its production rate is not expected to dominate over the lighter configurations, the much harder event topology makes the successive cuts extremely effective. In fact, the intermediate cut-flow reaches very large benchmark-level sensitivities before the final mass window, and the final benchmark-centered signal-region definition still preserves
\begin{equation}
\mathcal Z = 5.61,\ 10.24,\ 17.73
\end{equation}
at
\begin{equation}
L=300,\ 1000,\ 3000~\fb,
\end{equation}
respectively. This is the clearest example in the present study of hard kinematics compensating part of the rate suppression associated with large masses. 

As in the neutral channel, however, the results reported here should be interpreted strictly as benchmark-level statistical sensitivities within the present reconstruction and cut-based framework. They are not discovery claims, and they do not include systematic uncertainties or a more refined reconstruction strategy. Even with these caveats, the heavy charged-Higgs mode emerges as one of the strongest phenomenological outcomes of the analysis and as the most robust charged-scalar channel among those considered in this work.

\section{Conclusions}
\label{sec:conclusions}

We have presented a benchmark-level collider study of three flavor-violating Higgs channels in the Type-III Two-Higgs-Doublet Model at \(\sqrt s=14\) TeV:
\(pp\to H\to t\bar c \ (\bar t c)\),
\(pp\to H^\pm\to c\bar b \ (\bar c b)\), and
\(pp\to H^\pm\to t\bar b \ (\bar t b)\).
All channels were analyzed within a common simulation chain based on \textsc{MadGraph5\_aMC@NLO}, \textsc{Pythia8}, \textsc{Delphes3}, and \textsc{MadAnalysis5}, and their significance was quantified through the statistical estimator \(S/\sqrt{S+B}\) at 300, 1000, and 3000~\(\fb\).

For the neutral channel \(H\to t\bar c \ (\bar t c)\), the most favorable benchmark in the selected scan is
\((m_H,\tan\beta,\chi_{tc})=(180~\text{GeV},0.1,10)\),
with \(\mathcal Z=5.81,\ 10.61,\ 18.38\) at 300, 1000, and 3000~\(\fb\), respectively. More broadly, the neutral mode remains statistically relevant over an extended mass interval, and the most competitive points are concentrated in the low-\(\tan\beta\) region. Within the present comparative study, this channel constitutes one of the clearest probes of flavor-violating Higgs interactions in the up-quark sector.

For the light charged channel \(H^\pm\to c\bar b\ (\bar c b)\), the benchmark
\((m_{H^\pm},\tan\beta,\chi_{cb})=(110~\text{GeV},5,5)\),
selected as the most competitive configuration from the full scan, reaches \(\mathcal Z=6.18\) at 3000~\(\fb\) after an improved dijet-based selection. This result shows that the channel cannot be regarded as uniformly noncompetitive. At the same time, the final background remains extremely large, and the dominant samples are affected by large event weights, so the corresponding significance must be interpreted with caution. In this sense, the light charged mode is best viewed as a channel whose apparent reach is strongly dependent on the analysis strategy and which requires improved background statistics and a more refined treatment before any stronger conclusion can be drawn.

For the heavy charged channel \(H^\pm\to t\bar b \ (\bar t b)\), several benchmark configurations remain statistically sizable over the intermediate-mass region between 300 and 500 GeV, especially for large \(\chi\). The strongest charged-Higgs benchmark in the present comparison is
\((m_{H^\pm},\tan\beta,\chi)=(1000~\text{GeV},0.1,0.5)\),
which reaches \(\mathcal Z=5.61,\ 10.24,\ 17.73\) at 300, 1000, and 3000~\(\fb\), respectively. This result illustrates that, within the present cut-based framework, the loss in inclusive rate at high mass can be substantially compensated by the strong kinematic discrimination of the final state. Among the two charged-Higgs modes studied here, this is the most robust one.

Taken together, the results establish a clear phenomenological ordering among the three channels considered. The neutral and heavy charged modes emerge as the most robust benchmark probes within the present analysis, while the light charged mode remains significantly more fragile and more sensitive to the details of the event selection. This conclusion is stable at the benchmark level across the luminosity range studied here.

The analysis also clarifies the role of the final mass-window selections. In all three channels, the benchmark-centered \(\mjj\) window is not introduced merely as an optimization device; rather, it defines the mass interval in which the surviving signal support is localized after the preceding kinematic cuts. Consequently, even when the final mass window reduces the raw value of \(S/\sqrt{S+B}\), it still provides the physically relevant measure of how much of the selected significance is associated with the Higgs-mass hypothesis under study.

The results reported here should be interpreted strictly as benchmark-level statistical sensitivities. They do not include systematic uncertainties, nor do they replace a full experimental analysis with optimized reconstruction, improved control of combinatorial ambiguities, and larger Monte Carlo samples for the dominant backgrounds. These limitations are particularly important for the light charged channel, but they are relevant for the full comparison. With this caveat clearly stated, the present study nevertheless identifies the neutral and heavy charged flavor-violating modes as the most promising directions within the benchmark regions explored in the \ddHmIII.

Further work should include the incorporation of systematic effects, improved background statistics, and more refined reconstruction strategies, especially for the dijet-based charged-Higgs analysis. It would also be worthwhile to extend the present benchmark-level comparison toward more targeted search strategies adapted to the specific topology of each channel. Even at its current level, however, the analysis provides a coherent collider comparison of three complementary flavor-violating Higgs signatures and delineates the benchmark regions in which the \ddHmIII\ remains most competitive at the \LHC\ and \HL.

\section*{Data Availability Statement}
The data supporting the findings of this study are available from the corresponding author upon reasonable request. The benchmark yields and cut-flow results relevant to the main conclusions are provided in the manuscript and its Appendix.

\begin{acknowledgments}
We thank M. A. Arroyo-Ure\~na for valuable discussions and technical support. We are also grateful to A. Herrera Aguilar, F. Pacheco V\'azquez, and E.L. Sadurn\'i Hern\'andez for valuable comments and technical support during the evaluation of this work. M.L.F.P. acknowledges the support of CONAHCYT (now SECIHTI) and BUAP. A.R.S. acknowledges financial support from SNII. This work was supported by the Vicerrector\'ia de Investigaci\'on y Estudios de Posgrado of BUAP. 
\end{acknowledgments}

\clearpage
\onecolumngrid
\appendix

\section{General benchmark yields and final statistical significances}
\label{app:yields}

\renewcommand{\arraystretch}{1.0}
\setlength{\LTleft}{1pt}
\setlength{\LTright}{1pt}

\small
\begin{longtable}{|c|c|c|c|c|c|r|c|}
\caption{General benchmark yields and final statistical significances for the selected configurations of the neutral and charged Higgs channels. Only central values are shown.}
\label{tab:general_yields}\\
\hline
\textbf{Process} & \textbf{$m_H$ [GeV]} & \textbf{tan$\beta$} & \textbf{$\chi$} & \textbf{$L$} & \textbf{$S$} & \textbf{$B$} & \textbf{$Z$} \\
\hline
\endfirsthead

\multicolumn{8}{c}%
{{\bfseries Table \thetable\ continued from previous page}}\\
\hline
\textbf{Process} & \textbf{$m_H$ [GeV]} & \textbf{tan$\beta$} & \textbf{$\chi$} & \textbf{$L$} & \textbf{$S$} & \textbf{$B$} & \textbf{$Z$} \\
\hline
\endhead

\hline
\multicolumn{8}{r}{{Continued on next page}}\\
\endfoot

\hline
\endlastfoot

\(pp \to H \to t\bar c \ (\bar t c)\) & 180  & 0.1 & 10  & 300  & 71712   & 152142526    & 5.8126 \\
\(pp \to H \to t\bar c \ (\bar t c)\) & 180  & 0.1 & 10  & 1000 & 239042  & 507141756    & 10.6123 \\
\(pp \to H \to t\bar c \ (\bar t c)\) & 180  & 0.1 & 10  & 3000 & 717127  & 1521425269   & 18.3810 \\
\hline
\(pp \to H \to t\bar c \ (\bar t c)\) & 200  & 1   & 5   & 300  & 21468   & 145485226    & 1.780 \\
\(pp \to H \to t\bar c \ (\bar t c)\) & 200  & 1   & 5   & 1000 & 71563   & 484950754    & 3.249 \\
\(pp \to H \to t\bar c \ (\bar t c)\) & 200  & 1   & 5   & 3000 & 214689  & 1454852263   & 5.628 \\
\hline
\(pp \to H \to t\bar c \ (\bar t c)\) & 220  & 0.1 & 1   & 300  & 26271   & 140234164    & 2.2183 \\
\(pp \to H \to t\bar c \ (\bar t c)\) & 220  & 0.1 & 1   & 1000 & 87571   & 467447216    & 4.0500 \\
\(pp \to H \to t\bar c \ (\bar t c)\) & 220  & 0.1 & 1   & 3000 & 262713  & 1402341649   & 7.0148 \\
\hline
\(pp \to H \to t\bar c \ (\bar t c)\) & 240  & 0.1 & 1   & 300  & 48948   & 133214365    & 4.2402 \\
\(pp \to H \to t\bar c \ (\bar t c)\) & 240  & 0.1 & 1   & 1000 & 163163  & 444047886    & 7.7415 \\
\(pp \to H \to t\bar c \ (\bar t c)\) & 240  & 0.1 & 1   & 3000 & 489489  & 1332143658   & 13.4087 \\
\hline
\(pp \to H \to t\bar c \ (\bar t c)\) & 280  & 0.1 & 0.5 & 300  & 24319   & 116746889    & 2.250 \\
\(pp \to H \to t\bar c \ (\bar t c)\) & 280  & 0.1 & 0.5 & 1000 & 81063   & 389156296    & 4.109 \\
\(pp \to H \to t\bar c \ (\bar t c)\) & 280  & 0.1 & 0.5 & 3000 & 243190  & 1167468890   & 7.117 \\
\hline
\(pp \to H \to t\bar c \ (\bar t c)\) & 300  & 0.1 & 0.5 & 300  & 29657   & 113879759    & 2.7788 \\
\(pp \to H \to t\bar c \ (\bar t c)\) & 300  & 0.1 & 0.5 & 1000 & 98859   & 379599196    & 5.0734 \\
\(pp \to H \to t\bar c \ (\bar t c)\) & 300  & 0.1 & 0.5 & 3000 & 296577  & 1138797590   & 8.7874 \\
\hline
\(pp \to H \to t\bar c \ (\bar t c)\) & 320  & 0.1 & 0.5 & 300  & 35140   & 109326509    & 3.3603 \\
\(pp \to H \to t\bar c \ (\bar t c)\) & 320  & 0.1 & 0.5 & 1000 & 117136  & 364421698    & 6.1351 \\
\(pp \to H \to t\bar c \ (\bar t c)\) & 320  & 0.1 & 0.5 & 3000 & 351408  & 1093265095   & 10.6262 \\
\hline
\(pp \to H \to t\bar c \ (\bar t c)\) & 340  & 0.1 & 0.5 & 300  & 41295   & 102616531    & 4.0758 \\
\(pp \to H \to t\bar c \ (\bar t c)\) & 340  & 0.1 & 0.5 & 1000 & 137651  & 342055106    & 7.4413 \\
\(pp \to H \to t\bar c \ (\bar t c)\) & 340  & 0.1 & 0.5 & 3000 & 412955  & 1026165318   & 12.8887 \\
\hline
\(pp \to H^\pm \to c\bar b \ (\bar c b)\) & 110  & 5   & 5   & 300  & 119440  & 3731832723   & 1.9552 \\
\(pp \to H^\pm \to c\bar b \ (\bar c b)\) & 110  & 5   & 5   & 1000 & 398133  & 12439442411  & 3.5696 \\
\(pp \to H^\pm \to c\bar b \ (\bar c b)\) & 110  & 5   & 5   & 3000 & 1194399 & 37318327233  & 6.1827 \\
\hline
\(pp \to H^\pm \to t\bar b \ (\bar t b)\) & 300  & 1   & 10  & 300  & 347236  & 12069030146  & 3.16070 \\
\(pp \to H^\pm \to t\bar b \ (\bar t b)\) & 300  & 1   & 10  & 1000 & 1157456 & 40230100489  & 5.77062 \\
\(pp \to H^\pm \to t\bar b \ (\bar t b)\) & 300  & 1   & 10  & 3000 & 3472368 & 120690301468 & 9.99501 \\
\hline
\(pp \to H^\pm \to t\bar b \ (\bar t b)\) & 300  & 20  & 10  & 300  & 410230  & 12069030146  & 3.73408 \\
\(pp \to H^\pm \to t\bar b \ (\bar t b)\) & 300  & 20  & 10  & 1000 & 1367433 & 40230100489  & 6.81747 \\
\(pp \to H^\pm \to t\bar b \ (\bar t b)\) & 300  & 20  & 10  & 3000 & 4102301 & 120690301468 & 11.80821 \\
\hline
\(pp \to H^\pm \to t\bar b \ (\bar t b)\) & 350  & 1   & 10  & 300  & 273010  & 7124434140   & 3.23442 \\
\(pp \to H^\pm \to t\bar b \ (\bar t b)\) & 350  & 1   & 10  & 1000 & 910035  & 23748113800  & 5.90521 \\
\(pp \to H^\pm \to t\bar b \ (\bar t b)\) & 350  & 1   & 10  & 3000 & 2730107 & 71244341401  & 10.22813 \\
\hline
\(pp \to H^\pm \to t\bar b \ (\bar t b)\) & 350  & 20  & 10  & 300  & 349382  & 7124434140   & 4.13919 \\
\(pp \to H^\pm \to t\bar b \ (\bar t b)\) & 350  & 20  & 10  & 1000 & 1164609 & 23748113800  & 7.55710 \\
\(pp \to H^\pm \to t\bar b \ (\bar t b)\) & 350  & 20  & 10  & 3000 & 3493829 & 71244341401  & 13.08928 \\
\hline
\(pp \to H^\pm \to t\bar b \ (\bar t b)\) & 400  & 1   & 10  & 300  & 235557  & 4668289292   & 3.44753 \\
\(pp \to H^\pm \to t\bar b \ (\bar t b)\) & 400  & 1   & 10  & 1000 & 785193  & 15560964306  & 6.29430 \\
\(pp \to H^\pm \to t\bar b \ (\bar t b)\) & 400  & 1   & 10  & 3000 & 2355579 & 46682892920  & 10.90205 \\
\hline
\(pp \to H^\pm \to t\bar b \ (\bar t b)\) & 400  & 20  & 10  & 300  & 282558  & 4668289292   & 4.13539 \\
\(pp \to H^\pm \to t\bar b \ (\bar t b)\) & 400  & 20  & 10  & 1000 & 941861  & 15560964306  & 7.55015 \\
\(pp \to H^\pm \to t\bar b \ (\bar t b)\) & 400  & 20  & 10  & 3000 & 2825585 & 46682892920  & 13.07725 \\
\hline
\(pp \to H^\pm \to t\bar b \ (\bar t b)\) & 450  & 1   & 10  & 300  & 188535  & 3121106482   & 3.37463 \\
\(pp \to H^\pm \to t\bar b \ (\bar t b)\) & 450  & 1   & 10  & 1000 & 628452  & 10403688274  & 6.16121 \\
\(pp \to H^\pm \to t\bar b \ (\bar t b)\) & 450  & 1   & 10  & 3000 & 1885358 & 31211064823  & 10.67153 \\
\hline
\(pp \to H^\pm \to t\bar b \ (\bar t b)\) & 450  & 20  & 10  & 300  & 224446  & 3121106482   & 4.01738 \\
\(pp \to H^\pm \to t\bar b \ (\bar t b)\) & 450  & 20  & 10  & 1000 & 748155  & 10403688274  & 7.33471 \\
\(pp \to H^\pm \to t\bar b \ (\bar t b)\) & 450  & 20  & 10  & 3000 & 2244466 & 31211064823  & 12.70408 \\
\hline
\(pp \to H^\pm \to t\bar b \ (\bar t b)\) & 500  & 20  & 10  & 300  & 157752  & 1536267218   & 4.02458 \\
\(pp \to H^\pm \to t\bar b \ (\bar t b)\) & 500  & 20  & 10  & 1000 & 525841  & 5120890728   & 7.34785 \\
\(pp \to H^\pm \to t\bar b \ (\bar t b)\) & 500  & 20  & 10  & 3000 & 1577525 & 15362672185  & 12.72684 \\
\hline
\(pp \to H^\pm \to t\bar b \ (\bar t b)\) & 1000 & 0.1 & 0.5 & 300  & 86057   & 235435362    & 5.6076 \\
\(pp \to H^\pm \to t\bar b \ (\bar t b)\) & 1000 & 0.1 & 0.5 & 1000 & 286858  & 784784541    & 10.2380 \\
\(pp \to H^\pm \to t\bar b \ (\bar t b)\) & 1000 & 0.1 & 0.5 & 3000 & 860576  & 2354353623   & 17.7327 \\
\end{longtable}
\normalsize

\clearpage
\twocolumngrid

\section{Additional benchmark distributions}
\label{app:benchmark_plots}

In this appendix we collect supplementary kinematic distributions corresponding to the competitive benchmark points selected in the main analysis. Their role is illustrative: even for statistically competitive configurations, the signal does not generally emerge as a visually distinctive structure over the total background after the last cut.

\subsection{Neutral channel: $pp\to H\to t\bar{c}+\bar{t}c$}

\begin{figure}[H]
    \centering
    \includegraphics[width=\linewidth]{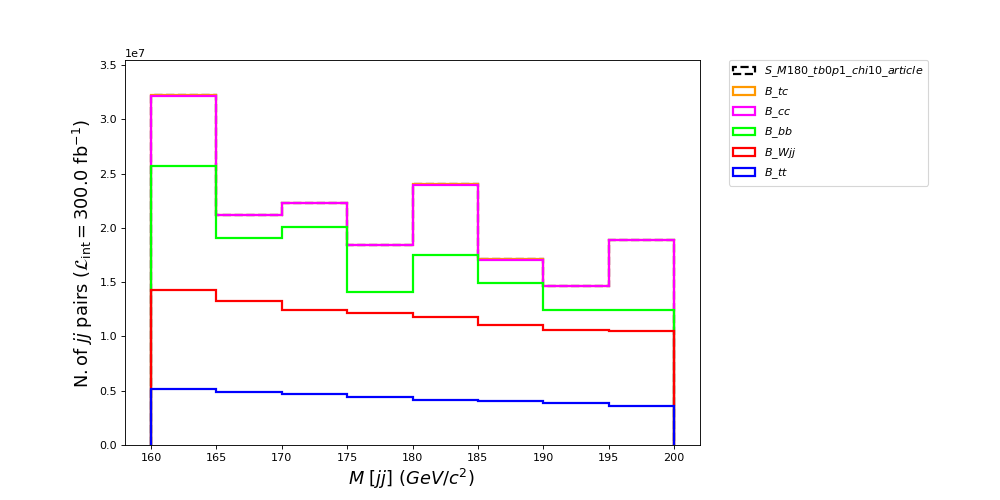}
    \caption{Neutral benchmark with $m_H=180$ GeV, $\tan\beta=0.1$, and $\chi_{tc}=10$.}
    \label{fig:app_Htc_180_0p1_10}
\end{figure}

\begin{figure}[H]
    \centering
    \includegraphics[width=\linewidth]{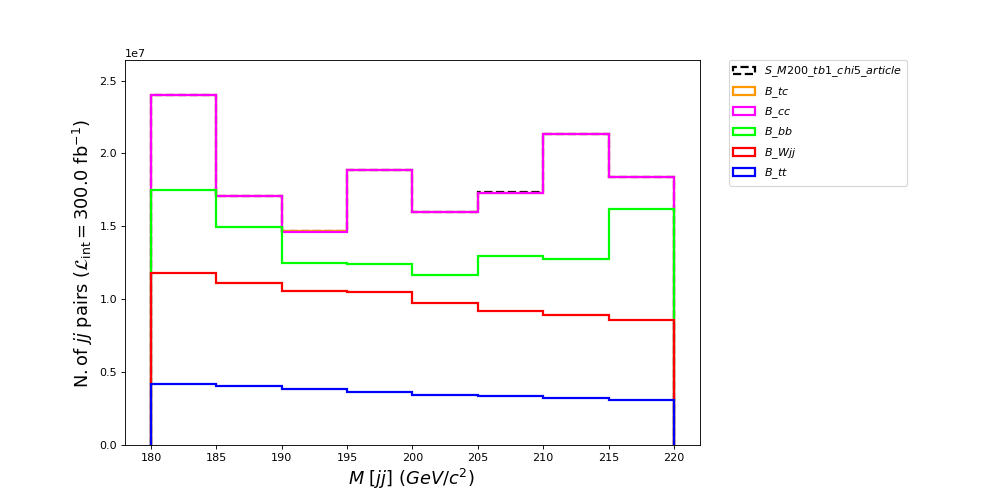}
    \caption{Neutral benchmark with $m_H=200$ GeV, $\tan\beta=1$, and $\chi_{tc}=5$.}
    \label{fig:app_Htc_200_1_5}
\end{figure}

\begin{figure}[H]
    \centering
    \includegraphics[width=\linewidth]{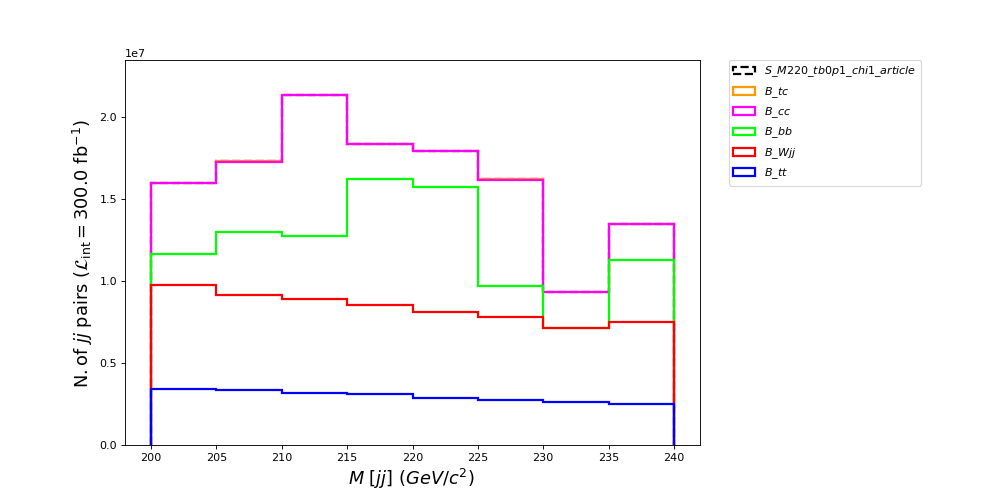}
    \caption{Neutral benchmark with $m_H=220$ GeV, $\tan\beta=0.1$, and $\chi_{tc}=1$.}
    \label{fig:app_Htc_220_0p1_1}
\end{figure}

\begin{figure}[H]
    \centering
    \includegraphics[width=\linewidth]{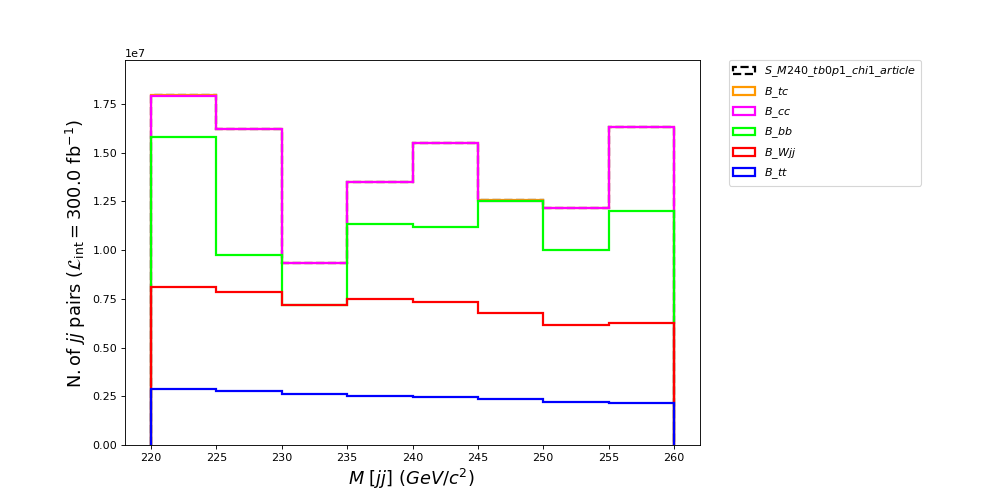}
    \caption{Neutral benchmark with $m_H=240$ GeV, $\tan\beta=0.1$, and $\chi_{tc}=1$.}
    \label{fig:app_Htc_240_0p1_1}
\end{figure}

\begin{figure}[H]
    \centering
    \includegraphics[width=\linewidth]{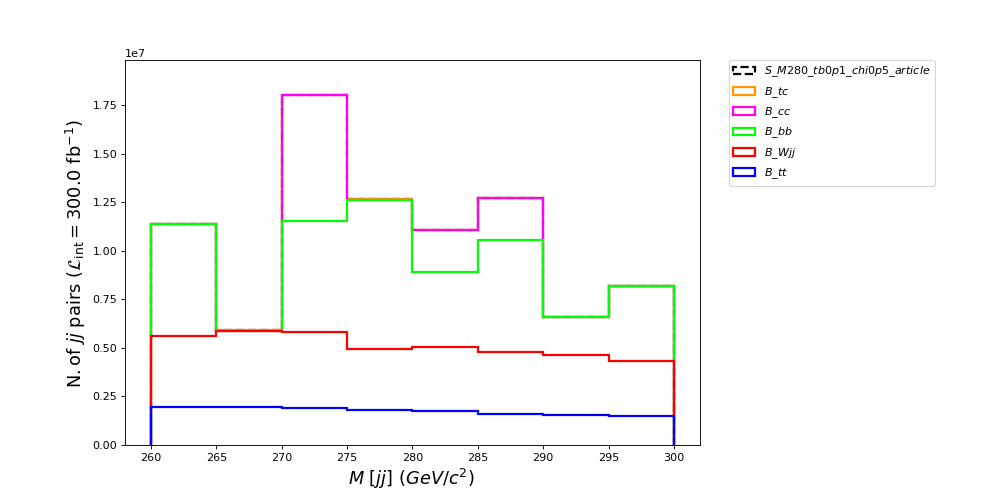}
    \caption{Neutral benchmark with $m_H=280$ GeV, $\tan\beta=0.1$, and $\chi_{tc}=0.5$.}
    \label{fig:app_Htc_280_0p1_0p5}
\end{figure}

\begin{figure}[H]
    \centering
    \includegraphics[width=\linewidth]{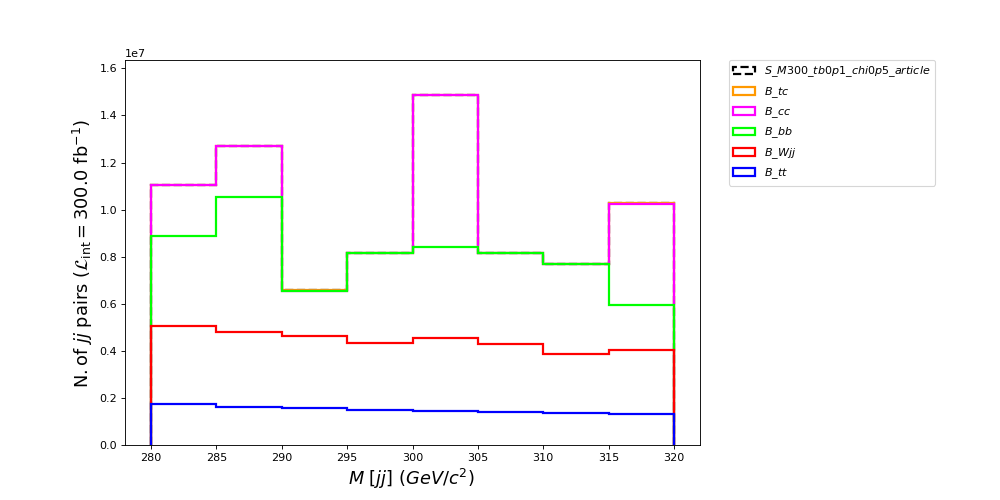}
    \caption{Neutral benchmark with $m_H=300$ GeV, $\tan\beta=0.1$, and $\chi_{tc}=0.5$.}
    \label{fig:app_Htc_300_0p1_0p5}
\end{figure}

\begin{figure}[H]
    \centering
    \includegraphics[width=\linewidth]{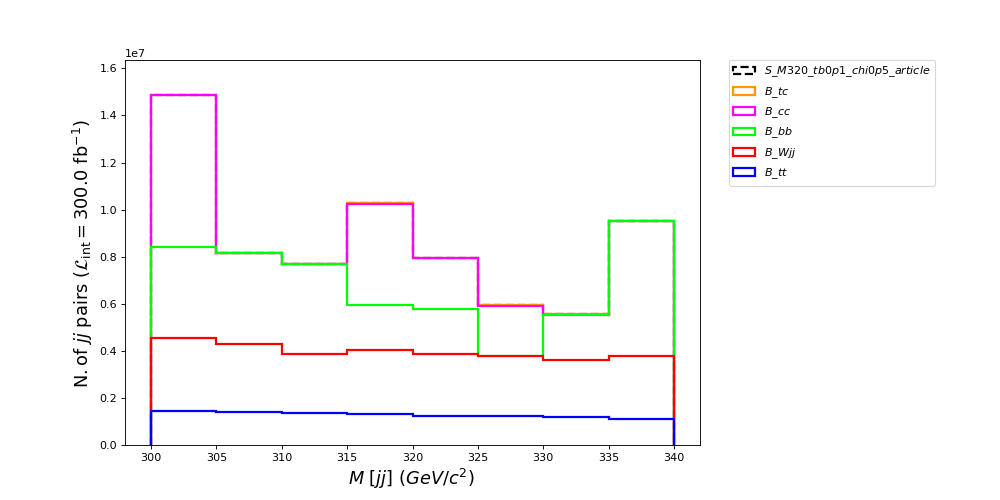}
    \caption{Neutral benchmark with $m_H=320$ GeV, $\tan\beta=0.1$, and $\chi_{tc}=0.5$.}
    \label{fig:app_Htc_320_0p1_0p5}
\end{figure}

\begin{figure}[H]
    \centering
    \includegraphics[width=\linewidth]{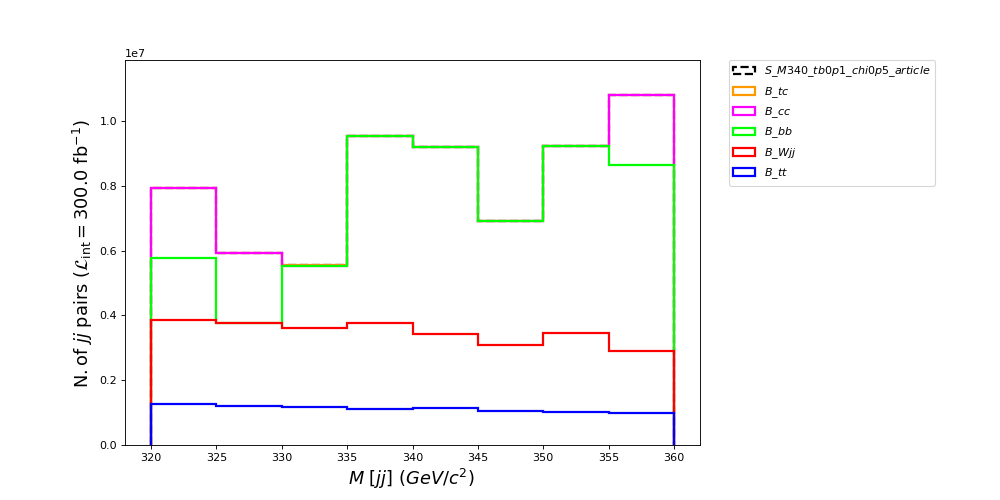}
    \caption{Neutral benchmark with $m_H=340$ GeV, $\tan\beta=0.1$, and $\chi_{tc}=0.5$.}
    \label{fig:app_Htc_340_0p1_0p5}
\end{figure}

\subsection{Light charged channel: $pp\to H^\pm\to c\bar{b}+\bar{c}b$}

\begin{figure}[H]
    \centering
    \includegraphics[width=\linewidth]{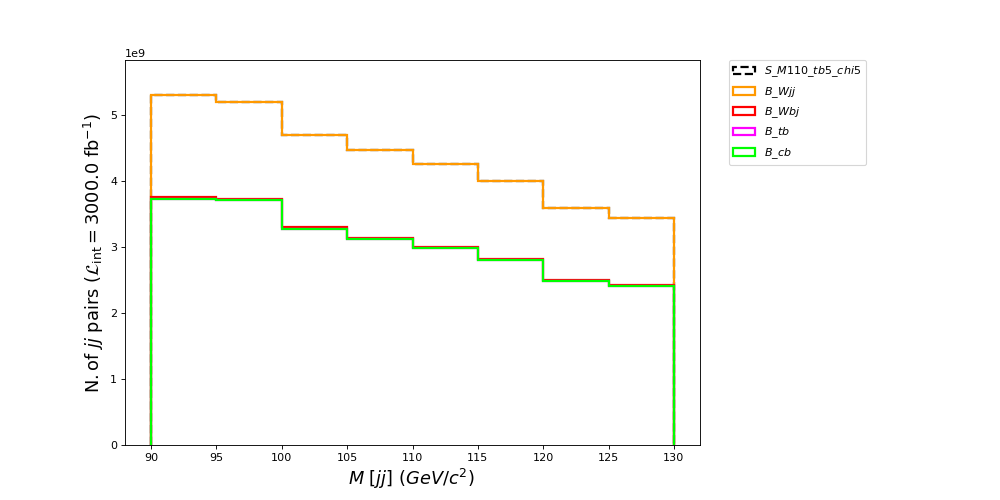}
    \caption{Improved light charged-Higgs benchmark with $m_{H^\pm}=110$ GeV, $\tan\beta=5$, and $\chi_{cb}=5$, showing the final dijet mass distribution in the selected $90<M(jj)<130$ GeV region.}
    \label{fig:app_Hpcb_110_5_5}
\end{figure}

\subsection{Heavy charged channel: $pp\to H^\pm\to t\bar{b}+\bar{t}b$}

\begin{figure}[H]
    \centering
    \includegraphics[width=\linewidth]{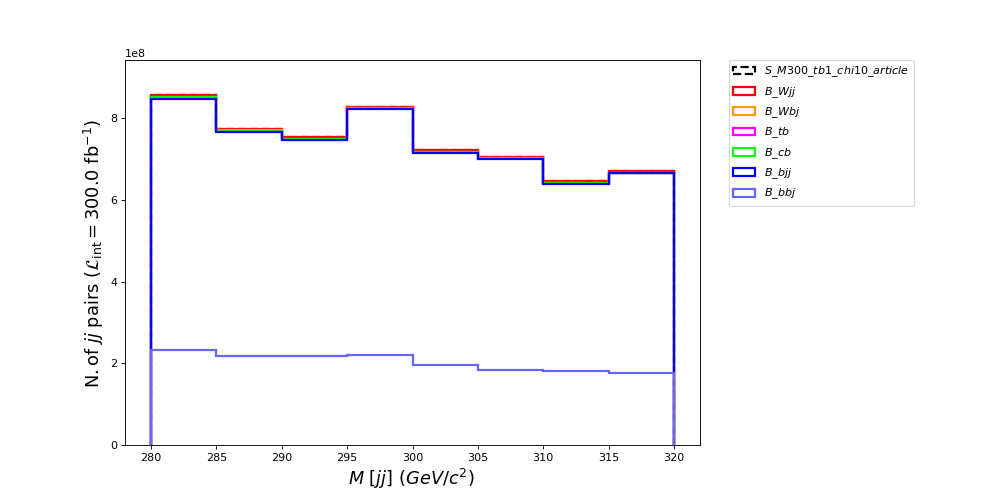}
    \caption{Charged benchmark with $m_{H^\pm}=300$ GeV, $\tan\beta=1$, and $\chi=10$.}
    \label{fig:app_Hptb_300_1_10}
\end{figure}

\begin{figure}[H]
    \centering
    \includegraphics[width=\linewidth]{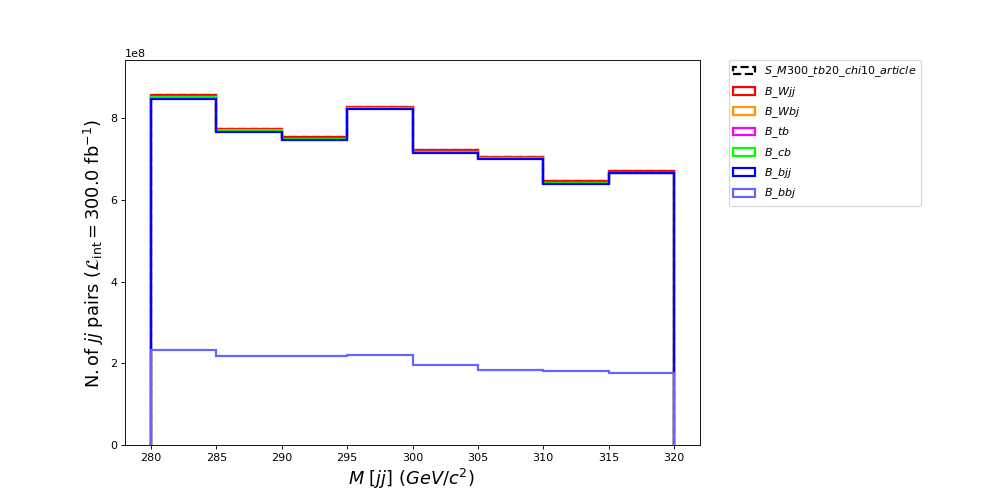}
    \caption{Charged benchmark with $m_{H^\pm}=300$ GeV, $\tan\beta=20$, and $\chi=10$.}
    \label{fig:app_Hptb_300_20_10}
\end{figure}

\begin{figure}[H]
    \centering
    \includegraphics[width=\linewidth]{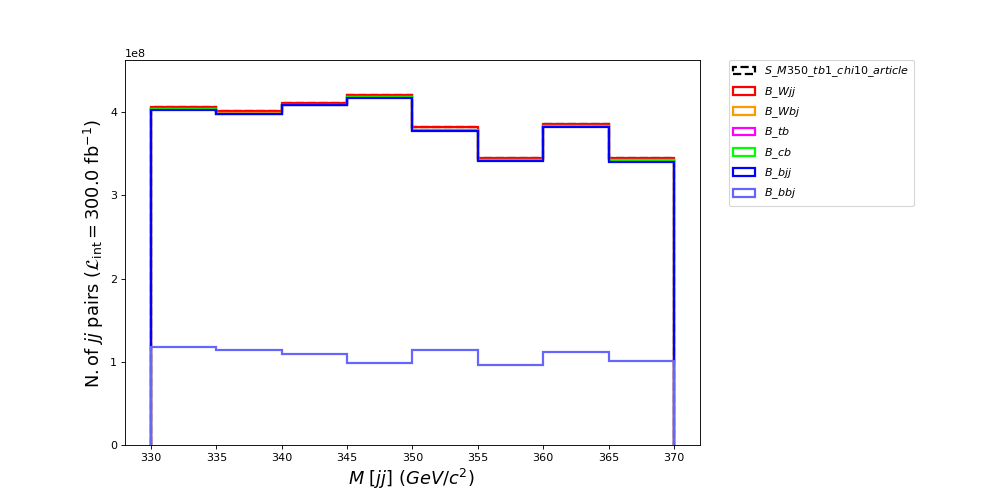}
    \caption{Charged benchmark with $m_{H^\pm}=350$ GeV, $\tan\beta=1$, and $\chi=10$.}
    \label{fig:app_Hptb_350_1_10}
\end{figure}

\begin{figure}[H]
    \centering
    \includegraphics[width=\linewidth]{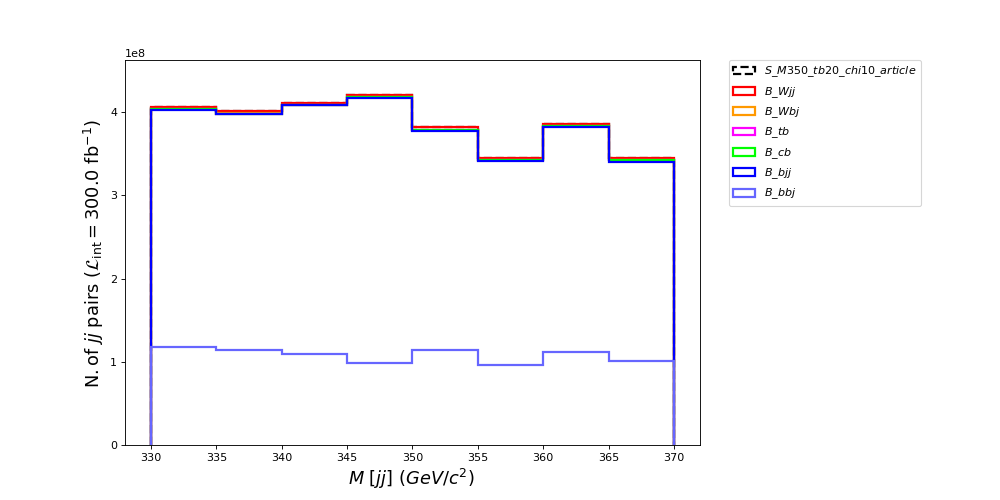}
    \caption{Charged benchmark with $m_{H^\pm}=350$ GeV, $\tan\beta=20$, and $\chi=10$.}
    \label{fig:app_Hptb_350_20_10}
\end{figure}

\begin{figure}[H]
    \centering
    \includegraphics[width=\linewidth]{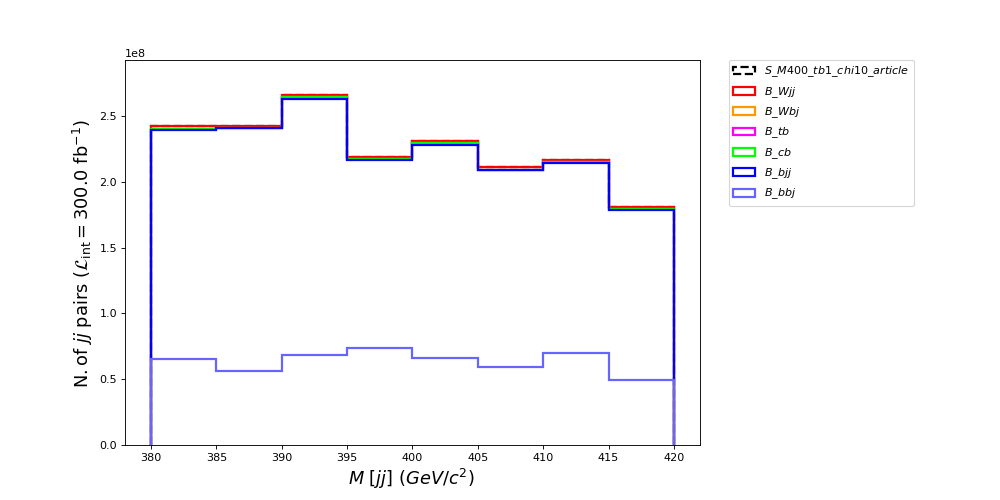}
    \caption{Charged benchmark with $m_{H^\pm}=400$ GeV, $\tan\beta=1$, and $\chi=10$.}
    \label{fig:app_Hptb_400_1_10}
\end{figure}

\begin{figure}[H]
    \centering
    \includegraphics[width=\linewidth]{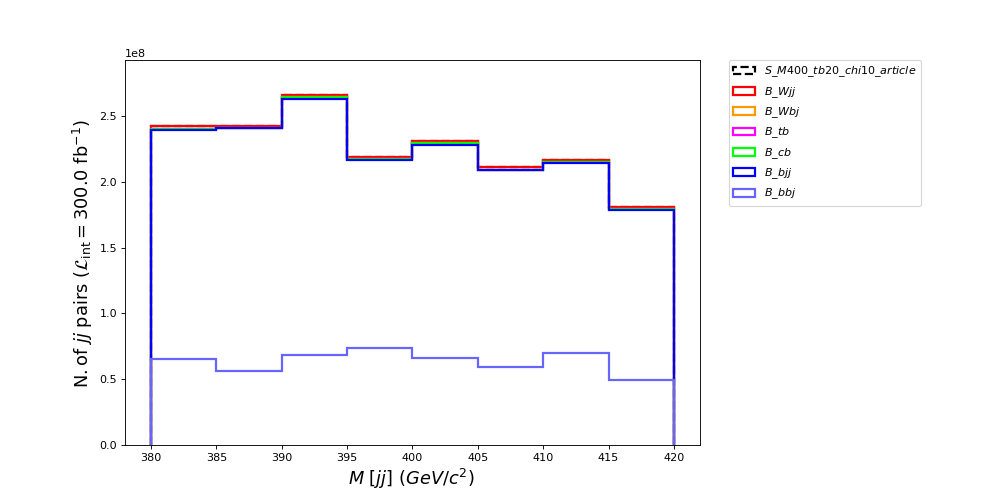}
    \caption{Charged benchmark with $m_{H^\pm}=400$ GeV, $\tan\beta=20$, and $\chi=10$.}
    \label{fig:app_Hptb_400_20_10}
\end{figure}

\begin{figure}[H]
    \centering
    \includegraphics[width=\linewidth]{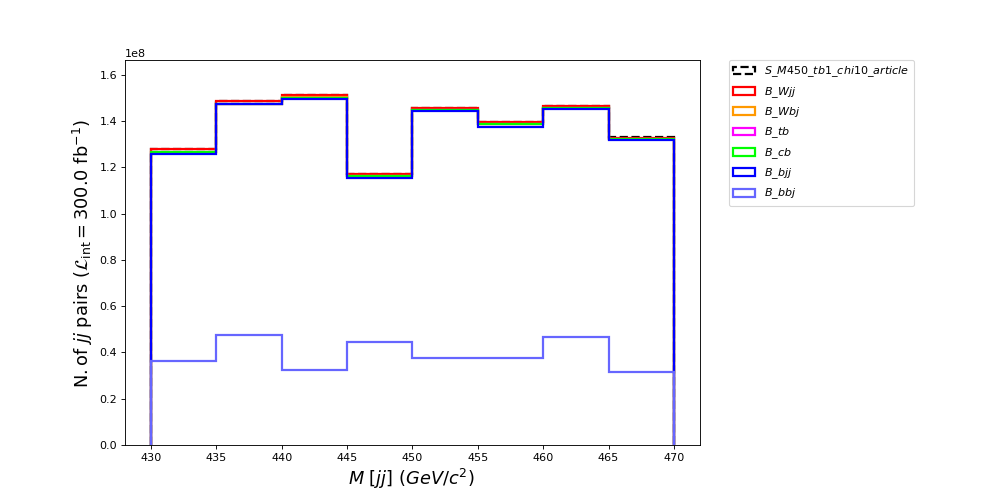}
    \caption{Charged benchmark with $m_{H^\pm}=450$ GeV, $\tan\beta=1$, and $\chi=10$.}
    \label{fig:app_Hptb_450_1_10}
\end{figure}

\begin{figure}[H]
    \centering
    \includegraphics[width=\linewidth]{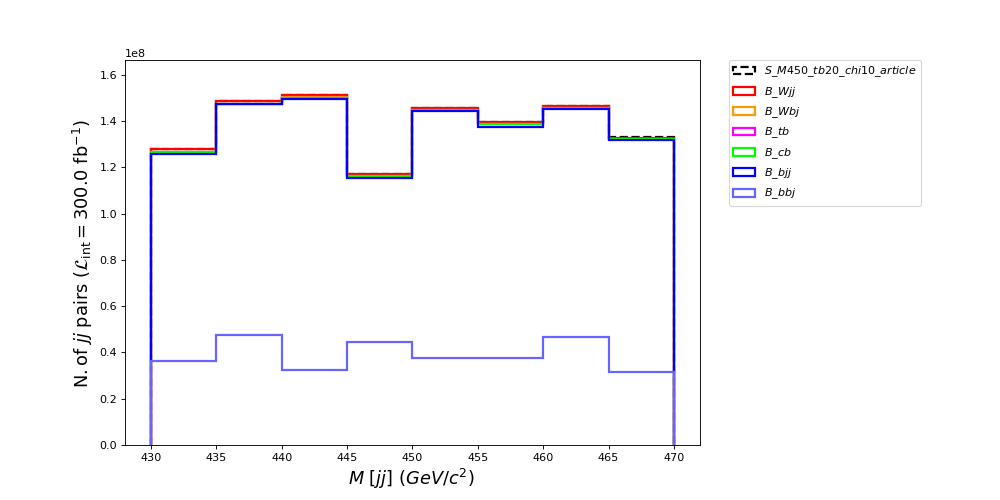}
    \caption{Charged benchmark with $m_{H^\pm}=450$ GeV, $\tan\beta=20$, and $\chi=10$.}
    \label{fig:app_Hptb_450_20_10}
\end{figure}

\begin{figure}[H]
    \centering
    \includegraphics[width=\linewidth]{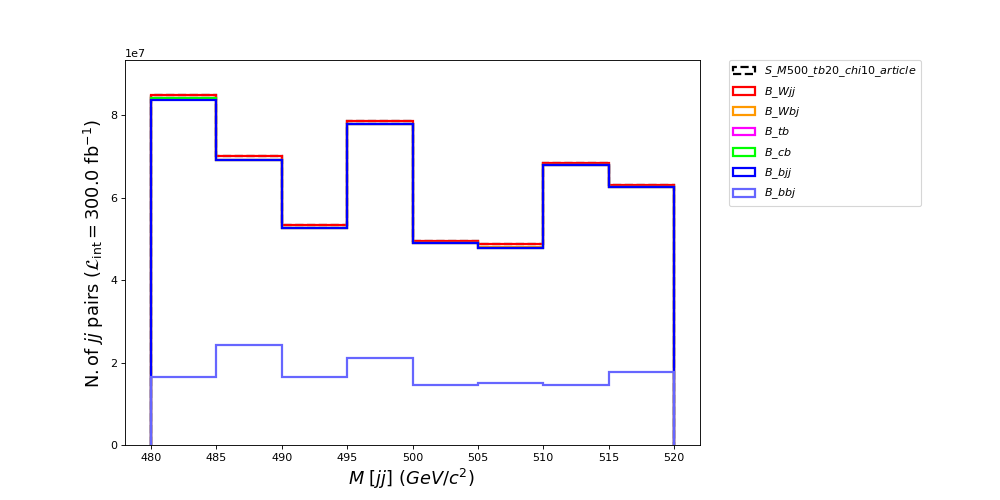}
    \caption{Charged benchmark with $m_{H^\pm}=500$ GeV, $\tan\beta=20$, and $\chi=10$.}
    \label{fig:app_Hptb_500_20_10}
\end{figure}

\begin{figure}[H]
    \centering
    \includegraphics[width=\linewidth]{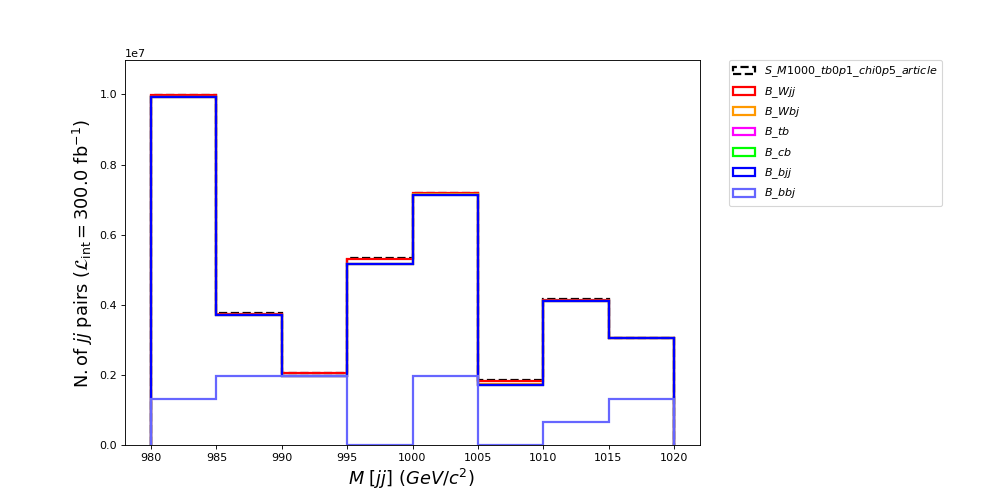}
    \caption{Charged benchmark with $m_{H^\pm}=1000$ GeV, $\tan\beta=0.1$, and $\chi=0.5$.}
    \label{fig:app_Hptb_1000_0p1_0p5}
\end{figure}
\bibliographystyle{apsrev4-2}
\bibliography{References}

@article{Glashow1961,
  author       = {Glashow, Sheldon L.},
  title        = {Partial-Symmetries of Weak Interactions},
  journal      = {Nuclear Physics},
  volume       = {22},
  year         = {1961},
  pages        = {579--588},
  doi          = {10.1016/0029-5582(61)90469-2}
}

@article{Weinberg1967,
  author       = {Weinberg, Steven},
  title        = {A Model of Leptons},
  journal      = {Physical Review Letters},
  volume       = {19},
  year         = {1967},
  pages        = {1264--1266},
  doi          = {10.1103/PhysRevLett.19.1264}
}

@incollection{Salam1968,
  author       = {Salam, Abdus},
  title        = {Weak and Electromagnetic Interactions},
  booktitle    = {Elementary Particle Theory: Relativistic Groups and Analyticity},
  editor       = {Svartholm, Nils},
  series       = {Nobel Symposium},
  number       = {8},
  publisher    = {Almqvist and Wiksell},
  address      = {Stockholm},
  year         = {1968},
  pages        = {367--377}
}

@article{HiggsDiscovery_ATLAS,
  author       = {{ATLAS Collaboration}},
  title        = {Observation of a New Particle in the Search for the Standard Model Higgs Boson with the {ATLAS} Detector at the {LHC}},
  journal      = {Physics Letters B},
  volume       = {716},
  year         = {2012},
  pages        = {1--29},
  doi          = {10.1016/j.physletb.2012.08.020},
  archivePrefix= {arXiv},
  eprint       = {1207.7214},
  primaryClass = {hep-ex}
}

@article{HiggsDiscovery_CMS,
  author       = {{CMS Collaboration}},
  title        = {Observation of a New Boson at a Mass of 125 {GeV} with the {CMS} Experiment at the {LHC}},
  journal      = {Physics Letters B},
  volume       = {716},
  year         = {2012},
  pages        = {30--61},
  doi          = {10.1016/j.physletb.2012.08.021},
  archivePrefix= {arXiv},
  eprint       = {1207.7235},
  primaryClass = {hep-ex}
}

@article{Branco2012_2HDMReview,
  author       = {Branco, G. C. and Ferreira, P. M. and Lavoura, L. and Rebelo, M. N. and Sher, Marc and Silva, Jo{\~a}o P.},
  title        = {Theory and Phenomenology of Two-Higgs-Doublet Models},
  journal      = {Physics Reports},
  volume       = {516},
  number       = {1-2},
  year         = {2012},
  pages        = {1--102},
  doi          = {10.1016/j.physrep.2012.02.002},
  archivePrefix= {arXiv},
  eprint       = {1106.0034},
  primaryClass = {hep-ph}
}

@article{ChengSher1987,
  author       = {Cheng, Ta-Pei and Sher, Marc},
  title        = {Mass Matrix Ansatz and Flavor Nonconservation in Models with Multiple Higgs Doublets},
  journal      = {Physical Review D},
  volume       = {35},
  year         = {1987},
  pages        = {3484--3491},
  doi          = {10.1103/PhysRevD.35.3484}
}

@book{HiggsHuntersGuide,
  author       = {Gunion, John F. and Haber, Howard E. and Kane, Gordon L. and Dawson, Sally},
  title        = {The Higgs Hunter's Guide},
  series       = {Frontiers in Physics},
  volume       = {80},
  publisher    = {CRC Press},
  address      = {Boca Raton},
  year         = {2000},
  isbn         = {9780738203058}
}

@article{Atwood1997_TypeIII,
  author       = {Atwood, David and Reina, Laura and Soni, Amarjit},
  title        = {Phenomenology of Two Higgs Doublet Models with Flavor-Changing Neutral Currents},
  journal      = {Physical Review D},
  volume       = {55},
  year         = {1997},
  pages        = {3156--3176},
  doi          = {10.1103/PhysRevD.55.3156},
  archivePrefix= {arXiv},
  eprint       = {hep-ph/9609279},
  primaryClass = {hep-ph}
}

@article{DiazCruz2004_TypeIII,
  author       = {D{\'i}az-Cruz, J. Lorenzo and Noriega-Papaqui, R. and Rosado, A.},
  title        = {Mass Matrix Ansatz and Lepton Flavor Violation in the Two-Higgs-Doublet Model Type III},
  journal      = {Physical Review D},
  volume       = {69},
  year         = {2004},
  pages        = {095002},
  doi          = {10.1103/PhysRevD.69.095002},
  archivePrefix= {arXiv},
  eprint       = {hep-ph/0401194},
  primaryClass = {hep-ph}
}

@article{HernandezSanchez2012,
  author       = {Hern{\'a}ndez-S{\'a}nchez, J. and L{\'o}pez-Lozano, L. and Noriega-Papaqui, R. and Rosado, A.},
  title        = {Couplings of Quarks in the Partially Aligned 2HDM with a Four-Zero Texture Yukawa Matrix},
  journal      = {Physical Review D},
  volume       = {85},
  year         = {2012},
  pages        = {071301},
  doi          = {10.1103/PhysRevD.85.071301},
  archivePrefix= {arXiv},
  eprint       = {1106.5035},
  primaryClass = {hep-ph}
}

@article{HernandezSanchez2015,
  author       = {F{\'e}lix-Beltr{\'a}n, O. and Gonz{\'a}lez-Canales, F. and Hern{\'a}ndez-S{\'a}nchez, J. and Moretti, S. and Noriega-Papaqui, R. and Rosado, A.},
  title        = {Analysis of the Quark Sector in the 2HDM with a Four-Zero Yukawa Texture},
  journal      = {Physics Letters B},
  volume       = {742},
  year         = {2015},
  pages        = {347--354},
  doi          = {10.1016/j.physletb.2015.02.003}
}

@article{ArroyoUrena2016_THDMtx,
  author       = {Arroyo-Ure{\~n}a, Marco A. and D{\'i}az-Cruz, J. Lorenzo and D{\'i}az, Enrique and Orduz-Ducuara, Javier A.},
  title        = {Flavor Violating Higgs Signals in the Texturized Two-Higgs Doublet Model ({THDM-Tx})},
  journal      = {Chinese Physics C},
  volume       = {40},
  number       = {12},
  year         = {2016},
  pages        = {123103},
  doi          = {10.1088/1674-1137/40/12/123103},
  archivePrefix= {arXiv},
  eprint       = {1306.2343},
  primaryClass = {hep-ph}
}

@article{ArroyoUrena2019_tch,
  author       = {Arroyo-Ure{\~n}a, Marco A. and Gait{\'a}n-Lozano, R. and Herrera-Chac{\'o}n, E. A. and Montes de Oca Y., J. H. and Valencia-P{\'e}rez, T. A.},
  title        = {Search for the $t \to ch$ Decay at Hadron Colliders},
  journal      = {Journal of High Energy Physics},
  volume       = {07},
  year         = {2019},
  pages        = {041},
  doi          = {10.1007/JHEP07(2019)041},
  archivePrefix= {arXiv},
  eprint       = {1903.02718},
  primaryClass = {hep-ph}
}

@article{GomezBock2024_FCNCtop,
  author       = {G{\'o}mez-Bock, M. and Gonzalez-Olivares, W. and Hentschinski, M. and Rosado Navarro, S.},
  title        = {Top Quark Production Through Flavor Violating Neutral Currents Within the 2HDM Type-III},
  journal      = {International Journal of Modern Physics A},
  volume       = {39},
  number       = {30},
  year         = {2024},
  pages        = {2450110},
  doi          = {10.1142/S0217751X24501100}
}

@article{HaberStal2015,
  author       = {Haber, Howard E. and St{\aa}l, Oscar},
  title        = {New {LHC} Benchmarks for the $\mathcal{CP}$-Conserving Two-Higgs-Doublet Model},
  journal      = {European Physical Journal C},
  volume       = {75},
  year         = {2015},
  pages        = {491},
  doi          = {10.1140/epjc/s10052-015-3697-x},
  archivePrefix= {arXiv},
  eprint       = {1507.04281},
  primaryClass = {hep-ph}
}

@article{Ivanov2017,
  author       = {Ivanov, Igor P.},
  title        = {Building and Testing Models with Extended Higgs Sectors},
  journal      = {Progress in Particle and Nuclear Physics},
  volume       = {95},
  year         = {2017},
  pages        = {160--208},
  doi          = {10.1016/j.ppnp.2017.03.001},
  archivePrefix= {arXiv},
  eprint       = {1702.03776},
  primaryClass = {hep-ph}
}

@article{HernandezSanchez2020_cbFusion,
  author       = {Hernandez-Sanchez, J. and Honorato, C. G. and Moretti, S. and Rosado-Navarro, S.},
  title        = {Charged Higgs Boson Production via $cb$-Fusion at the Large Hadron Collider},
  journal      = {Physical Review D},
  volume       = {102},
  year         = {2020},
  pages        = {055008},
  doi          = {10.1103/PhysRevD.102.055008},
  archivePrefix= {arXiv},
  eprint       = {2003.06263},
  primaryClass = {hep-ph}
}

@article{ArroyoUrena2025_Hpp,
  author       = {Arroyo-Ure{\~n}a, M. A. and Herrera-Chac{\'o}n, E. A. and Rosado Navarro, S. and Salazar, H.},
  title        = {Hunting for a Charged Higgs Boson Pair in Proton-Proton Collisions},
  journal      = {Physical Review D},
  volume       = {111},
  year         = {2025},
  pages        = {015023},
  doi          = {10.1103/PhysRevD.111.015023}
}

@article{Alwall2014_MadGraph5,
  author       = {Alwall, J. and Frederix, R. and Frixione, S. and Hirschi, V. and Maltoni, F. and Mattelaer, O. and Shao, H.-S. and Stelzer, T. and Torrielli, P. and Zaro, M.},
  title        = {The Automated Computation of Tree-Level and Next-to-Leading Order Differential Cross Sections, and Their Matching to Parton Shower Simulations},
  journal      = {Journal of High Energy Physics},
  volume       = {07},
  year         = {2014},
  pages        = {079},
  doi          = {10.1007/JHEP07(2014)079},
  archivePrefix= {arXiv},
  eprint       = {1405.0301},
  primaryClass = {hep-ph}
}

@article{Sjostrand2015_Pythia8,
  author       = {Sj{\"o}strand, Torbj{\"o}rn and Ask, Stefan and Christiansen, Jesper R. and Corke, Richard and Desai, Nishita and Ilten, Philip and Mrenna, Stephen and Prestel, Stefan and Rasmussen, Christine O. and Skands, Peter Z.},
  title        = {An Introduction to {PYTHIA} 8.2},
  journal      = {Computer Physics Communications},
  volume       = {191},
  year         = {2015},
  pages        = {159--177},
  doi          = {10.1016/j.cpc.2015.01.024},
  archivePrefix= {arXiv},
  eprint       = {1410.3012},
  primaryClass = {hep-ph}
}

@article{deFavereau2014_Delphes3,
  author       = {de Favereau, J. and Delaere, C. and Demin, P. and Giammanco, A. and Lema{\^i}tre, V. and Mertens, A. and Selvaggi, M.},
  title        = {{DELPHES} 3: A Modular Framework for Fast Simulation of a Generic Collider Experiment},
  journal      = {Journal of High Energy Physics},
  volume       = {02},
  year         = {2014},
  pages        = {057},
  doi          = {10.1007/JHEP02(2014)057},
  archivePrefix= {arXiv},
  eprint       = {1307.6346},
  primaryClass = {hep-ex}
}

@article{Conte2013_MA5,
  author       = {Conte, Eric and Fuks, Benjamin and Serret, Guillaume},
  title        = {{MadAnalysis} 5, a User-Friendly Framework for Collider Phenomenology},
  journal      = {Computer Physics Communications},
  volume       = {184},
  number       = {1},
  year         = {2013},
  pages        = {222--256},
  doi          = {10.1016/j.cpc.2012.09.009},
  archivePrefix= {arXiv},
  eprint       = {1206.1599},
  primaryClass = {hep-ph}
}

@article{Conte2014_MA5_update,
  author       = {Conte, Eric and Dumont, B{\'e}ranger and Fuks, Benjamin and Wymant, Chris},
  title        = {{MadAnalysis} 5 for New Physics Searches at the {LHC}},
  journal      = {European Physical Journal C},
  volume       = {74},
  year         = {2014},
  pages        = {3103},
  doi          = {10.1140/epjc/s10052-014-3103-0},
  archivePrefix= {arXiv},
  eprint       = {1405.3982},
  primaryClass = {hep-ph}
}

\end{document}